\documentclass[twocolumn,prb,amsmath,amssymb,floatfix,showpacs]{revtex4-2}

\usepackage{tikz}
\usetikzlibrary{positioning}
\usetikzlibrary{calc}

\usepackage{graphicx}
\usepackage[caption=false]{subfig}
\usepackage{ulem}

\usepackage{amsthm}
\usepackage{bm}
\usepackage{hyperref}

\newcommand\Tr{\mathrm{Tr}}

\newcommand\bbR{\mathbb{R}}
\newcommand\bfk{\mathbf{k}}
\newcommand\bfn{\mathbf{n}}
\newcommand\bfr{\mathbf{r}}

\newcommand\bfv{\mathbf{v}}
\newcommand\bfx{\mathbf{x}}
\newcommand\bfy{\mathbf{y}}

\newcommand\calG{\mathcal{G}}

\newcommand{\threej}[6]{\left( \begin{array}{@{}c@{\;}c@{\;}c@{}}
         #1 & #2 & #3 \\
         #4 & #5 & #6
         \end{array}\right)}

\newcommand{\isoscalar}[6]{\left( \begin{array}{@{}c@{\;}c@{\;}c@{\;}c@{}}
         #1 & #2 & | & #3 \\
         #4 & #5 & | & #6 
         \end{array}\right)}

\newcommand{\sixj}[6]{\left\{ \begin{array}{@{}c@{\;}c@{\;}c@{}}
         #1 & #2 & #3 \\
         #4 & #5 & #6
         \end{array}\right\}}

\begin{document}
\title{Parametrization of the Coulomb interaction matrix with point-group symmetry}

\author{Coraline Letouz\'e}
\email[Corresponding author: ] {coraline.letouze@sorbonne-universite.fr}
\author{Guillaume Radtke}
\author{Benjamin Lenz}
\author{Christian Brouder}
\affiliation{Sorbonne Universit\'e, Mus\'eum National d'Histoire Naturelle, 
UMR CNRS 7590, Institut de Min\'eralogie, de Physique des Mat\'eriaux et 
de Cosmochimie, IMPMC, 75005 Paris, France}

\date{\today}

\begin{abstract}
Coulomb integrals, i.e. matrix elements of bare or screened Coulomb 
interaction between one-electron orbitals, are fundamental objects in many 
approaches developed to tackle the challenging problem of calculating the electronic 
structure of strongly correlated materials. In this paper, Coulomb integrals are 
analyzed by considering both the point group symmetry of the site occupied by the atom 
in the crystal or molecule and the permutation symmetries of the orbitals in the integrals. 
In particular, the case where one-electron orbitals form the basis of a general (i.e. a real, 
complex or pseudo-complex) irreducible representation is considered. 
Explicit formulas are provided to calculate all integrals of the interaction tensor in terms
of a minimum set of independent ones.
The effect of a symmetry breaking is also investigated by describing Coulomb integrals 
of a group in terms of those of one of its subgroups. We develope the specific example
of $O(3)$ as the larger group which can therefore be used to quantify the deviation of a specific
system from the spherical symmetry.
Possible applications of the presented framework include the calculation of solid-state and molecular 
spectroscopies via multiplet techniques, dynamical mean-field theory or the GW approximation.
\end{abstract}
\maketitle

\section{Introduction}
Electronic correlations play a fundamental role in determining the properties of compounds
with partially filled $d$- or $f$-shells. Strong Coulomb interactions occuring between 
electrons occupying these localized orbitals are indeed among the most important parameters 
favouring, for instance, a particular ground-state symmetry of the ions. Thus, they determine 
their magnetic properties~\cite{Okimoto-21,Kumar-21}, induce metal-insulator 
transitions~\cite{Gebhard-97, Imada-98}, superconductivity or trigger long-range ordering 
phenomena involving either charge, orbital or spin degrees of freedom~\cite{Khomskii-14}. 

From a theoretical point of view, the explicit inclusion of local Coulomb interactions between 
correlated electrons beyond single-particle approaches often relies on the density-density 
approximation, where only dominant direct Coulomb and exchange terms are retained from 
the full tensor. These terms are then frequently averaged over the manifold of correlated orbitals, 
resulting in effective Hubbard $U$ and Hund's exchange $J_{\text{H}}$ parameters~\cite{vanderMarel-88}. 
The averaged parameters are commonly employed in standard implementations of the so-called LDA+$U$ 
method, where an effective single-particle approach based on the local density approximation is 
corrected in the manifold of correlated orbitals by on-site Hubbard and exchange 
terms~\cite{Cococcioni-12, Anisimov-91}. 
The orbitally averaged density-density approximation is also frequently employed in Green's 
function based many-body techniques, such as dynamical mean-field theory 
(DMFT)~\cite{Lichtenstein-98,Aichhorn-10,Hariki-17,Leonov-20}.

While this approach might be accurate enough to approximate the ground-state 
properties of materials in many cases, it is clearly not sufficient to provide a good 
description of the full multiplet structure accessible by many solid-state or molecular
spectroscopies, ranging from infrared or visible light optical absorption~\cite{Balhausen-62}, 
X-ray absorption or (non) resonant inelastic X-ray scattering to electron energy loss 
spectroscopy~\cite{DeGroot-08}.
In this case indeed, a full account of the Coulomb tensor within and between 
the correlated electronic shells involved in the excitation process is mandatory but 
theoretically and numerically very challenging.  For decades, this problem has been 
tackled by assuming that transition metal or rare-earth ions retained a dominant 
atomic-like character in the molecular or solid state and, therefore, that Coulomb interaction
could be handled within the spherical symmetry~\cite{Haverkort-12}. A great advantage 
of this approximation lies in the fact that only a very limited number of numerical parameters, 
known as Slater integrals or Slater-Condon parameters~\cite{Slater-29,CondonShortley}, 
need to be introduced to parametrize  the full tensor. For example, if one considers the case 
of $d$ electrons, the $5^4 = 625$ elements of the spin-independent Coulomb tensor can be 
expressed in terms of only three Slater integrals,  $F^0$, $F^2$ and $F^4$ and 
simple expressions such as $U = F^0$ and  $J_{\text{H}} = (F^2+ F^4)/14$ are obtained. 
Also when considering $d$ electrons within a perfectly cubic symmetry represented by
real wavefunctions, effective descriptions like the Kanamori form~\cite{Kanamori-1963},
which goes beyond density-density interactions, can be expressed in terms of
these three Slater integrals~\cite{Georges-2013}.

The validity of this approximation is, however, questionable for ions in solids or molecules 
where the local symmetry of the atomic site is reduced and a covalent interaction 
with the surrounding ligand atoms always occurs to a certain degree.
In this case, the inclusion of the resulting anisotropy of the interaction can lead to important corrections, for instance for the Fermi surface~\cite{Zhang-16,Sarvestani-18}.
Recent progresses in the first-principle calculation of screened Coulomb interactions within the constrained 
random phase approximation in solids~\cite{Aryasetiawan-04} indeed show numerically that, 
whereas Slater parametrization is fairly accurate for ions with very localized states and in 
highly symmetric environnements, larger deviations are expected 
when increasing the spatial extension of the orbitals, the covalent character of the 
interaction with the ligands or reducing the local site symmetry~\cite{Vaugier-12,Panda-17,vanRoekeghem-16}. 
In such cases, a proper analysis of the effective interaction tensor should be carried out 
by accounting explicitly for the local point group symmetry of the atomic site. In particular,
the central question of the number of independent parameters required to describe exactly
the entire tensor immediately arises. This is indeed of primary importance in the analysis of 
spectroscopic data, since this number is the maximum
 number of independent parameters
to fit, but also when investigating numerically  the properties of realistic model Hamiltonians 
accounting for the exact spatial symmetry of the system.

This problem was pioneered by Tanabe, Sugano and Kamimura~\cite{TanabeSugano}
in the early 1970's for the specific case of $d$-shell electrons in octahedral $O_h$ 
symmetry. It was recently extended by B{\"u}nemann and Gebhard to the case of
$d$- and $f$-shell electrons in $O_h$, $O$, $T_d$, $T_h$, $D_{6h}$ and $D_{4h}$ 
symmetries~\cite{Bunemann-17}. 
Iimura, Hirayama and Hoshino followed a  different route and
expressed the anisotropic Coulomb tensor in terms of multipole operators~\cite{Iimura-21}. 
A general theory dealing with any orbital in any group is, however, still missing 
and is therefore the main focus of the present paper. In particular, we  consider here
all types of irreducible representations (irreps) whereas previous works only focussed on
real  wavefunctions. Moreover, we provide general expressions 
for the independent Coulomb parameters as well as for any 
Coulomb integral on the interaction tensor in terms of these parameters.
Finally, we  study the effect of a symmetry breaking by comparing Coulomb integrals of a group with 
those of one of its subgroups.

We would like to underline the broad applicability of our approach.
Indeed, we make only two assumptions: (i) the (possibly screened) electron-electron interaction
$U(\bfr,\bfr')$ is symmetric
(i.e.
$U(\bfr,\bfr')=U(\bfr',\bfr)$)
and invariant under the 
operations of a crystal point group $G$;
(ii) the basis functions $\phi^{(\alpha)}_a(\bfr)$
transform as the basis elements of an irrep $\alpha$
of $G$~\cite{noteA}.
In particular, we do
not assume that the wave functions
$\varphi^{(\alpha)}_a(\bfr)$
entering the Coulomb integrals are built from
spherical harmonics of a specific $\ell$,
nor do they need to have the same radial part.
In addition, $U(\bfr,\bfr')$ can
also be frequency-dependent (corresponding to a dynamical interaction)
since the frequency $\omega$ does not enter in the following derivations.
This renders the framework applicable to dynamical interactions, which are for instance used in the context 
of the GW approximation~\cite{Hedin-65,Aryasetiawan-98}, extended DMFT \cite{Sun-02} 
or techniques combining both~\cite{Biermann-03,Biermann-14}.

The paper is organized as follows.  Section~\ref{sec:invar} starts with a 
discussion of the various  symmetries of Coulomb integrals for complex and 
real one-electron orbitals. More specifically, we consider the case of 
one-electron orbitals forming bases for irreducible representations 
of a crystal point group $G$. In section~\ref{sec:grinv}, we use the Clebsch-Gordan
coefficients of $G$ to define linear combinations of Coulomb integrals 
(called $G$-invariants) that are invariant under the action of the operations 
of $G$ and we show that all Coulomb integrals can be written in terms of 
these $G$-invariants. 
Section~\ref{sec:perminv} describes how permutation
symmetries can be taken into account to further reduce the number 
of independent integrals, which are now called (permutation)-symmetrized 
$G$-invariants. In this section, we give an explicit formula for 
calculating any Coulomb integral in terms of these symmetrized $G$-invariants
and we show, conversely, that symmetrized $G$-invariants can be calculated 
from the same number of well-chosen Coulomb integrals. 
Section~\ref{sec:subd} explores the important case of symmetry breaking 
by considering that $G$ is the subgroup of a
larger group $\calG$ and presents
the expression of the $G$-invariants in terms of $\calG$
invariants.
The example where $\calG$ is the infinite group $O(3)$ is detailed to illustrate the
calculations. In this case, the relation
between
$O(3)$-invariants (related to Slater integrals) 
{and $G$-invariants} can be used to quantify 
the deviation of 
the system from spherical symmetry. In section~\ref{sec:concl} finally, we 
present our conclusions as well as possible extensions of our work.

\section{Invariance of Coulomb integrals}
\label{sec:invar}

In solid-state and molecular physics the electron-electron interaction 
between orbitals $\varphi_a$, $\varphi_b$, $\varphi_c$, $\varphi_d$ is
described by Coulomb integrals defined as
\begin{align}
U_{abcd} &= \langle \varphi_a \varphi_b | U | \varphi_c \varphi_d \rangle \nonumber \\
&= \int d\bfr d\bfr' \varphi_a(\bfr)^\ast \varphi_b(\bfr')^\ast U(\bfr, \bfr') \varphi_d(\bfr') \varphi_c(\bfr) ,
\label{eq:integral}
\end{align}
where $U(\bfr,\bfr')$ is proportional to $1/|\bfr-\bfr'|$ for the
bare electron-electron interaction but can be much more complicated
if we consider screened Coulomb interactions as we do here.
We assume $U(\bfr,\bfr')$ to be real (otherwise consider its real and imaginary 
parts separately) and permutation symmetric in the sense 
$U(\bfr', \bfr)=U(\bfr, \bfr')$. Note that the spin degree of freedom is 
not considered in the present work.

In this paper we focus on two kinds of symmetries of $U_{abcd}$: 
(i) the on-site symmetry represented by a crystal point group $G$
(section \ref{sec:symmetries_pointgroup}) and (ii) the permutation
of the orbitals (section \ref{sec:symmetries_permutations}).

\subsection{Invariance under point symmetry operations \label{sec:symmetries_pointgroup}}

\paragraph{Action of a group.}
In an abstract way, the action of a group $G$ on a 
vector space $X$ is a
linear operation
that associates to each pair $(R,x) \in G \times X$ an element of $X$ denoted 
by $R \triangleright x$. This operation satisfies
\begin{itemize}
\item[(i)] $1 \triangleright x = x \quad \forall x \in X$, where $1$ is the
identity element of the group;
\item[(ii)] $\forall R, S \in G$ and $\forall x \in X$, $R \triangleright
(S \triangleright x) = (R S) \triangleright x$.
\end{itemize}
For example, if $X=\bbR^3$ and $G$ is a point symmetry group
defined by matrices $R$, then $R\triangleright \bfr=R\bfr$.

\paragraph{Action of a group on functions.}
In molecular or solid-state physics, we deal with orbitals 
or (wave)functions $\varphi$, which are functions of $\bfr$.
The action of  the symmetry operation $S$ on $\varphi$ is
a new function $\varphi_S$ of $\bfr$ defined by
\begin{eqnarray*}
(S\triangleright \varphi)(\bfr) &=& \varphi_S(\bfr)=\varphi(S^{-1}\bfr),
\end{eqnarray*}
where the argument of $\varphi$ (originally denoted by $\bfr$) is replaced by $S^{-1}\bfr$ in $\varphi$.
The presence of $S^{-1}\bfr$ instead of $S\bfr$ is required by the axioms of 
an action. Indeed
\begin{eqnarray*}
\big(R\triangleright(S\triangleright\varphi)\big)(\bfr) &=& 
(R\triangleright\varphi_S)(\bfr)=  
\varphi_S(R^{-1}\bfr)
\\&=&
\varphi\big(S^{-1}(R^{-1}\bfr)\big)
=\varphi((RS)^{-1}\bfr) 
\\&=&
\big((RS)\triangleright\varphi\big)(\bfr)
=\varphi_{RS}(\bfr).
\end{eqnarray*}
We can now describe the action of a symmetry operation on
a Coulomb integral by its action on the orbitals (Schr\"odinger
representation)
\begin{eqnarray*}
R\triangleright U_{abcd} &=& \int d\bfr d\bfr' \varphi_a(R^{-1}\bfr)^\ast 
\varphi_b(R^{-1}\bfr')^\ast U(\bfr, \bfr') 
\\&&
\times \varphi_d(R^{-1}\bfr') \varphi_c(R^{-1}\bfr),
\end{eqnarray*}
which can be transformed into an action on $U(\bfr,\bfr')$ (Heisenberg representation) 
by a change of variable
\begin{eqnarray*}
R\triangleright U_{abcd} = \int d\bfr d\bfr' \varphi_a(\bfr)^\ast \varphi_b(\bfr')^\ast U(R\bfr, R\bfr') 
\varphi_d(\bfr') \varphi_c(\bfr),
\end{eqnarray*}
where we used the fact that symmetry operations preserve volumes
(i.e. $d(R\bfr)=d\bfr$). 
Thus, Coulomb integrals are invariant under the operations of the
group $G$ if 
$U$ is invariant under the operations
of $G$:
$U(R\bfr, R\bfr')=U(\bfr, \bfr')$ for all $R$ of $G$.
\textit{We assume this to hold in the present paper}.
As a consequence, the symmetry of the system
is expressed by the property of  the Coulomb integrals
\begin{equation}
  R \triangleright U_{abcd} = U_{abcd},
  \label{eq:invariance}
\end{equation}
for all $R$ of $G$.

\paragraph{Group representation.}
A group representation is a set of 
unitary matrices $\Gamma(R)$, one for each element $R$ of the group, 
which satisfies $\Gamma(R) \Gamma(S) = \Gamma(RS)$ for any two elements 
$R$ and $S$ in $G$. If $d$ is the dimension of the matrices,
a basis of the carrier space of this representation is a set of $d$ orbitals $\varphi_a, \varphi_b, \ldots$ 
such that
\begin{eqnarray}
R \triangleright \varphi_a &=& \sum_b \Gamma_{ba}(R) \varphi_b .
\label{eq:action_on_wavefunction}
\end{eqnarray}
The order of the indices of $\Gamma$ might appear surprising at first glance, 
but  is actually required~\cite{LudwigFalter} to satisfy the second property 
of the action
\begin{eqnarray*}
R\triangleright(S\triangleright\varphi_a)
&=& R \triangleright \big( \sum_b \Gamma_{ba}(S) \varphi_b \big) = 
\sum_b \Gamma_{ba}(S) (R \triangleright \varphi_b) \\
&=& \sum_{bc} \Gamma_{ba}(S) \Gamma_{cb}(R) \varphi_c = 
\sum_c (\Gamma(R)\Gamma(S))_{ca} \varphi_c \\
&=& \sum_c (\Gamma(RS))_{ca} \varphi_c = (RS) \triangleright \varphi_a ,
\end{eqnarray*}
where we used the linearity of the action in the first line, then that 
$\Gamma(R) \Gamma(S) = \Gamma(RS)$ between the second and third lines.

\paragraph{Irreducible representations.}
We are particulary interested in irreducible representations (irreps) which are 
representations that cannot be decomposed into smaller representations. 
In this paper, we assume that the orbitals transform as irreps (denoted by 
$\alpha$, $\beta$, $\gamma$ and $\delta$) of $G$. For each of these 
representations, for instance $\alpha$, let $\{ \varphi_a^{(\alpha)}\}$, 
$a=1,\dots,\dim \alpha$, be a basis of this representation and 
$\{ \Gamma^{(\alpha)} (R), \ \forall R \in G\}$ be its representation matrices.
We denote the Coulomb integral on the basis of the irreps as
\begin{eqnarray*}
U_{abcd}^{(\alpha\beta\gamma\delta)} &=&
\langle \varphi^{(\alpha)}_a\varphi^{(\beta)}_b | U | \varphi^{(\gamma)}_c \varphi^{(\delta)}_d\rangle.
\end{eqnarray*}
The action of $R$ on $U_{abcd}^{(\alpha\beta\gamma\delta)}$
can now be described by representation matrices
\begin{eqnarray}
R \triangleright U_{abcd}^{(\alpha\beta\gamma\delta)} &=&
\sum_{a'b'c'd'} \Gamma^{(\alpha)}_{a'a}(R)^* 
\Gamma^{(\beta)}_{b'b}(R)^* \Gamma^{(\gamma)}_{c'c}(R) \Gamma^{(\delta)}_{d'd}(R)
\nonumber\\&& \times U_{a'b'c'd'}^{(\alpha\beta\gamma\delta)}.
\label{I1234transf}
\end{eqnarray}

There are three types of irreps, in the sense of the Frobenius-Schur
 criterion~\cite{Frobenius-06}:
\begin{itemize}
\item[(a)] real irreps, for which we can find real representation matrices;
\item[(b)] pseudo-real (or quaternionic) irreps, for which $\Gamma(R)$ and 
$\Gamma(R)^*$  are equivalent in the sense that they are related by a 
similarity transformation, but they are not all equivalent to real representation 
matrices;
\item[(c)] complex irreps, for which $\Gamma(R)$ and $\Gamma(R)^*$ are 
not equivalent, meaning that they are associated to different irreps; an example 
of which is given by $e^{\pm i\bfk \cdot \bfr}$ for the translation group.
\end{itemize}

The distinction between real and non-real (i.e. quasi-real or complex) irreps is 
crucial. Indeed, in the case of real representations, we can choose real representation 
matrices and also real-valued basis functions $\{\varphi_a^{(\alpha)}\}$ of the carrier 
space of the irrep. Please note that even real representations can be represented by 
complex-valued matrices, as it is the case for instance of the real representation 
$E_g$ for the group $O_h$ in Altmann and Herzig's tables~\cite{Altmann}. 
However, in the case of non-real representations, the representation matrices and 
the basis functions cannot be chosen real-valued.

For the 
groups $O_h$, $O$, $T_d$, $D_{6h}$ and $D_{4h}$ investigated by B{\"u}nemann 
and Gebhard~\cite{Bunemann-17}, all irreps are real. For the last group $T_h$ studied 
in that article, there are four complex one-dimensional representations 
$({}^1E_g, {}^2E_g, {}^1E_u, {}^2E_u)$. They can be grouped into pairs to become 
two-dimensional real  representations,
which are however no longer irreducible.

\subsection{Invariance under some permutations \label{sec:symmetries_permutations}}

We now describe  additional symmetries of the Coulomb integrals, related to the
permutation of $\bfr$ and $\bfr'$ and the complex conjugation in 
Eq. \eqref{eq:integral}. The permutation symmetries differ between real- and 
complex-valued orbitals.

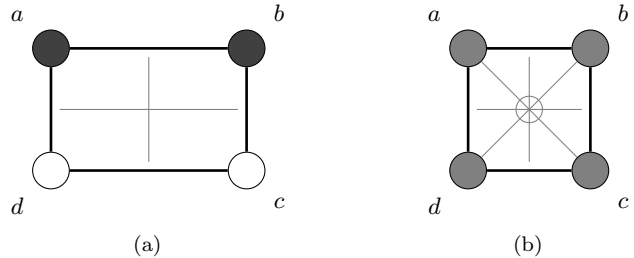
\begin{figure}
 \centering
 \def\length{5em}
 \def\bigside{8em}
 \def\smallside{5em}
 \def\border{0.5em}

 \subfloat[\label{fig:permutations_nonreal}]{
  \centering
  \begin{tikzpicture}
    \node[fill=black!75, draw=black, circle, minimum size=1.5em] (a) at (0, 0) {};
    \node[fill=black!75, draw=black, circle, minimum size=1.5em] (b) at (\bigside, 0)  {};
    \node[fill=white, draw=black, circle, minimum size=1.5em] (c) at (\bigside, -\smallside) {};
    \node[fill=white, draw=black, circle, minimum size=1.5em] (d) at (0, -\smallside) {};

    \node[above left = 0.1 of a] (alabel) {$a$};
    \node[above right = 0.1 of b] (blabel) {$b$};
    \node[below right = 0.1 of c] (clabel) {$c$};
    \node[below left = 0.1 of d] (dlabel) {$d$};

    \node (ad) at ($(a)!0.5!(d)$) {};
    \node (ab) at ($(a)!0.5!(b)$) {};
    \node (bc) at ($(b)!0.5!(c)$) {};
    \node (cd) at ($(c)!0.5!(d)$) {};

    \draw [line width= 1] (a) to (b);
    \draw [line width= 1] (b) to (c);
    \draw [line width= 1] (c) to (d);
    \draw [line width= 1] (d) to (a);

    \draw [color=gray] (ab) to (cd);
    \draw [color=gray] (ad) to (bc);
  \end{tikzpicture}
}
\hfill
\subfloat[\label{fig:permutations_real}]{
  \centering
  \begin{tikzpicture}
    \node [fill=gray, draw=black,circle, minimum size=1.5em] (a) at (0, 0) {};
    \node  [fill=gray, draw=black,circle, minimum size=1.5em] (b) at (\length, 0) {};
    \node  [fill=gray, draw=black,circle, minimum size=1.5em] (c) at (\length, -\length) {};
    \node  [fill=gray, draw=black,circle, minimum size=1.5em] (d) at (0, -\length) {};

    \node[above left = 0.1 of a] (alabel) {$a$};
    \node[above right = 0.1 of b] (blabel) {$b$};
    \node[below right = 0.1 of c] (clabel) {$c$};
    \node[below left = 0.1 of d] (dlabel) {$d$};

    \node (ad) at (0, -\length/2) {};
    \node (ab) at (\length/2, 0) {};
    \node (center) at (\length/2, -\length/2) {};
    \node (bc) at (\length, -\length/2) {};
    \node (cd) at (\length/2, -\length) {};

    \draw [line width= 1] (a) to (b);
    \draw [line width= 1] (b) to (c);
    \draw [line width= 1] (c) to (d);
    \draw [line width= 1] (d) to (a);

    \draw [color = gray] (a) to (c);
    \draw [color = gray] (b) to (d);
    \draw [color = gray] (ab) to (cd);
    \draw [color = gray] (ad) to (bc);
    \filldraw[fill=none, color = gray] (center) circle (5pt);
  \end{tikzpicture}
}
\caption{Geometric representation of the permutation symmetries of $U_{abcd}$. 
The vertices are labelled clockwise with the orbital indices $a$, $b$, $c$ and $d$.   
(a) The complex conjugation of non-real orbitals  is graphically represented 
as a transformation of black vertices
into white ones (and vice versa). The permutations
form the group $m'm2'$ which describes the symmetries of a rectangle with 2 white and 2 black
vertices. (b) The permutations of real orbitals form the group $D_4$ which describes
the symmetries of a square.}
\end{figure}

\paragraph{Non-real representations.}
In the (more general) case of non-real representations, we assume 
complex-valued orbitals $\varphi_a$. Interchanging $\bfr$ and $\bfr'$ in the 
integral definition of $U_{abcd}$ (Eq.~\eqref{eq:integral}) gives $U_{abcd}=U_{badc}$, 
and taking its complex conjugate yields $U_{abcd}^*=U_{cdab}$. As a consequence, 
we obtain an equality between four Coulomb integrals
\begin{equation}
U_{abcd} = U_{badc}= U_{cdab}^*= U_{dcba}^*.
\label{eq:permutations_nonreal}
\end{equation}
This can be seen as the invariance of the Coulomb integrals under the action of an 
additional group $G_P$ of four elements $\{p_1, p_2, p_3, p_4 \}$. Its action on 
$U_{abcd}$ is defined by
\begin{eqnarray*}
p_1 \triangleright U_{abcd} &=& U_{abcd},
\quad p_2 \triangleright U_{abcd} =U_{badc}, \\
p_3 \triangleright U_{abcd} &=& U_{cdab}^*,
\quad p_4 \triangleright U_{abcd} = U_{dcba}^*.
\end{eqnarray*}
and is represented graphically on Fig.~(\ref{fig:permutations_nonreal}).

Due to the complex conjugation in its action, $G_P$ is a magnetic group, of which the Coulomb integrals
form a corepresentation, as defined by Wigner~\cite{Wigner}.

More precisely, the permutation group $G_P$ is isomorphic
to the Shubnikov group of the third kind $m'm2'$. It has two unitary 
operations $E$ and $\sigma_y$, and two anti-unitary operations $C_{2z}$ 
and $\sigma_x$~\cite{Bradley-10}.

\paragraph{Real representations.}
In the case of real representations, we assume real-valued orbitals. 
The previous equality (Eq. \eqref{eq:permutations_nonreal}) still holds 
with $U_{abcd} = U_{cdba} =  U_{badc}  =  U_{dcba}$. In addition, the 
two orbitals $\varphi_a(\bfr)$ and $\varphi_c(\bfr)$, as well as 
$\varphi_b(\bfr')$ and $\varphi_d(\bfr')$, now play an equivalent role 
and can be interchanged, yielding equalities like $U_{abcd} = U_{cbad}$. 
We therefore get equalities between eight Coulomb integrals~\cite{Bunemann-17}
\begin{eqnarray}
 && U_{abcd} = U_{cdba} =  U_{badc}  =  U_{dcba}  \nonumber \\
&=& U_{adcb} = U_{cbad} = U_{bcda}  =  U_{dabc}.
\label{eq:permutations_real}
\end{eqnarray}
Again, these equalities can be interpreted as the invariance of the Coulomb 
integral under the action of a group $G_P = D_4$ of eight permutations 
(see Fig.~(\ref{fig:permutations_real})).

\section{Group invariants}
\label{sec:grinv}

We now introduce the central objects of the present paper. Since they are 
built to be invariant under the action of the considered group, we call them group 
invariants and denote them $I$. 
As we shall see, they correspond to the eigenvalues of
the interaction matrix, on the basis of the irreps of the group.

As a starting point, we consider the point group only, leaving the permutation 
groups to the next section.

\subsection{Clebsch-Gordan coefficients}

For any finite (or compact) group $G$, the tensor product  (also known as the 
direct product, or the Kronecker product~\cite{LudwigFalter}) of two irreps can 
be written as a direct sum of irreps, called a Clebsch-Gordan series~\cite{CornwellI}
\begin{equation}
\alpha \otimes \beta = \eta_1 \oplus \dots \oplus \eta_k,
\end{equation}
where we assume that no irrep $\eta_i$ appears more than once. In other words, 
we assume tensor products of irreps to be multiplicity-free. 
This assumption can be relaxed without conceptual complication by using 
methods which are now standard in the literature~\cite{Derome-65,Derome-66,
Butler-75,Broek-78,Butler,Newmarch-83-2, Newmarch-83-3}, but this leads to a 
cluttering of indices in formulas, that we prefer to avoid. Additionally, this assumption 
is actually justified for crystal
point groups. Indeed, their product tables~\cite{Altmann} indicate 
that all crystal point groups except $T$ and $T_h$ are multiplicity-free. Even the exceptions
$T$ and $T_h$ satisfy the assumption in the broader sense that their doubly occuring 
irreps can be distinguished by their symmetric and anti-symmetric nature.

At the level of the matrix representations of the irreps, a unitary transformation brings their
tensor product into a sum of the irrep matrices. The coefficients of this unitary transformation
are called Clebsch-Gordan coefficients. More precisely, we define the Clebsch-Gordan 
coefficients  to be any set of complex numbers $(\alpha a \beta b| \eta e)$ solving the 
equation~\cite{LudwigFalter}
\begin{equation}
\Gamma_{a'a}^{(\alpha)}(R) \Gamma_{b'b}^{(\beta)}(R)
= \sum_{\eta ee'}
(\alpha a' \beta b'|\eta e')  \Gamma_{e'e}^{(\eta)}(R)
(\alpha a \beta b|\eta e)^* .
\label{eq:CGdefinition}
\end{equation}

A possible solution of this equation is given by Dirl's formula~\cite{Dirl-79}
\begin{eqnarray}
(\alpha a \beta b|\eta e) &=&
\frac{\sum_R \Gamma_{a a_0}^{(\alpha)}(R) \Gamma_{b b_0}^{(\beta)}(R) 
\big(\Gamma_{e e_0}^{(\eta)}(R)\big)^*} {\sqrt{N(a_0,b_0,e_0)}},
\label{eq:CGDirl}
\end{eqnarray}
with
\begin{eqnarray*}
N(a_0,b_0,e_0) &=& \frac{|G|}{\dim\eta}
\sum_R \Gamma_{a_0 a_0}^{(\alpha)}(R) \Gamma_{b_0 b_0}^{(\beta)}(R) 
\big(\Gamma_{e_0 e_0}^{(\eta)}(R)\big)^*,
\end{eqnarray*}
where $|G|$ is the order of $G$ (ie. the number of its elements) and, for each 
triple $(\alpha,\beta,\eta)$, three components $(a_0,b_0,e_0)$ are chosen so that 
$N(a_0,b_0,e_0)\not=0$. Such components exist if $\eta$ belongs to the tensor 
product of $\alpha \otimes \beta$ (see also~\cite{CornwellI}). These 
Clebsch-Gordan coefficients are a generalization of the ones used in angular 
momentum theory. They satisfy orthogonality relations
\begin{eqnarray}
\sum_{ab} (\alpha a\beta b|\eta e)^* (\alpha a \beta b|\phi f) &=& \delta_{\eta\phi} \delta_{ef},
\label{eq:CGortho1}\\
\sum_{\eta e} (\alpha a\beta b|\eta e)^* (\alpha a' \beta b'|\eta e) &=& \delta_{aa'} \delta_{bb'},
\label{eq:CGortho2}
\end{eqnarray}
due to the fact that they are elements of a unitary 
matrix~\cite{LudwigFalter}.

The definition of Clebsch-Gordan coefficients as any solution of Eq.~\eqref{eq:CGDirl} 
is inspired by Derome and Sharp~(see note \cite{note1}). Note that this definition 
does not fully specify the Clebsch-Gordan coefficients, since multiplying them by a 
phase depending on $\alpha$, $\beta$ and $\gamma$ transforms a solution into another 
one. Other approaches choose these phases carefully in order to maximize the symmetry 
of Clebsch-Gordan coefficients~\cite{Butler-75,Butler,Reid-82,Newmarch-83-2,
Newmarch-83-3,Piepho,Chen-18}. However, these phases depend on each group and 
the Clebsch-Gordan coefficients given by Dirl's formula (Eq.~\eqref{eq:CGDirl}) generally 
do not satisfy these symmetries. We chose to use the Clebsch-Gordan coefficients as defined by 
Derome and Sharp for another, crucial but technical reason; the interested reader is invited 
to read the note~\cite{note2}.

\subsection{Definition and properties of the point-group invariants \label{sec:definvariant}}

As a first step of the symmetry analysis of Coulomb integrals, we consider the submatrix 
$U^{(\alpha \beta \gamma \delta)}$ of elements $U_{abcd}^{(\alpha\beta\gamma\delta)}$ 
for a given quadruple of irreps $\sigma=(\alpha \beta \gamma \delta)$. We define the 
associated $G$-invariant as
\begin{align}
I^{(\alpha\beta\gamma\delta,\eta)}   =
\sum_{abcd e} \frac{(\alpha a\beta b|\eta e)^*  U_{abcd}^{(\alpha\beta\gamma\delta)}  
(\gamma c \delta d|\eta e)} {\dim\eta},
\label{eq:definvariant}
\end{align}
where $\eta$ belongs to the Clebsch-Gordan series of both $\alpha\otimes\beta$ and 
$\gamma\otimes\delta$.

Note that $I^{(\alpha\beta\gamma\delta,\eta)}$ is basis-independent. Indeed, let 
$P^{(\alpha)} = \sum_a | \varphi_a^{(\alpha)} \rangle \langle \varphi_a^{(\alpha)} |$ be the
projector onto a representation $\alpha$. There is a way to map
tensor products of states into sums of states so that
\begin{eqnarray*}
(\gamma c \delta d|\eta e) &=& (\langle \varphi_c^{(\gamma)} | 
\otimes \langle \varphi_d^{(\delta)} |)\ | \varphi_e^{(\eta)} \rangle, \\
(\alpha a \beta b|\eta e)^* &=& \langle \varphi_e^{(\eta)} |\ (|\varphi_a^{(\alpha)} \rangle 
\otimes | \varphi_b^{(\beta)} \rangle).
\end{eqnarray*}
Recalling that $U^{(\alpha \beta \gamma \delta)}_{abcd} = \langle \varphi_a^{(\alpha)} 
\varphi_b^{(\beta)}  | U | \varphi_c^{(\gamma)} \varphi_d^{(\delta)} \rangle$, we obtain
\begin{eqnarray*}
 I^{(\alpha\beta\gamma\delta,\eta)} &=& \frac{1}{\dim\eta}\mathrm{Tr}\Big(P^{(\eta)}
 (P^{(\alpha)}\otimes P^{(\beta)}) U
(P^{(\gamma)}\otimes P^{(\delta)})\Big),
\end{eqnarray*}
which is basis independent. This means that $G$-invariants can be compared even 
if the matrices of the irreps are different. However, if $(\gamma,\delta)\not=(\alpha,\beta)$ 
they can differ from one another by a phase due to the phase ambiguity of Clebsch-Gordan 
coefficients.

The second result is a consequence of Schur's lemma: if $U_{abcd}^{(\alpha\beta\gamma\delta)}$ 
is invariant under the action of $G$ (i.e. satisfies Eq.~\eqref{eq:invariance}), then the matrix 
$U_{abcd}^{(\alpha\beta\gamma\delta)}$ is diagonalized by the Clebsch-Gordan coefficients 
and its eigenvalues are $I^{(\alpha\beta\gamma\delta,\eta)}$
\begin{eqnarray}
\sum_{abcd} (\alpha a \beta b|\phi f)^* U_{abcd}^{(\alpha\beta\gamma\delta)}
(\gamma c \delta d|\eta e)   &=& 
 \delta_{\phi\eta} \delta_{ef} I^{(\alpha\beta\gamma\delta,\eta)}. \nonumber \\
\label{Schur}
\end{eqnarray}
As a consequence, any $U_{abcd}^{(\alpha\beta\gamma\delta)}$
can be written explicitly in terms of $G$-invariants
\begin{eqnarray}
U_{abcd}^{(\alpha\beta\gamma\delta)}  &=& 
\sum_{\eta e} (\alpha a \beta b|\eta e)   I^{(\alpha\beta\gamma\delta,\eta)} 
(\gamma c \delta d|\eta e)^* .
\label{eq:Ufrominvariant}
\end{eqnarray}
This shows that, if $\eta$ runs over the $n^{(\alpha \beta \gamma \delta)}$ irreps 
shared by $\alpha \otimes \beta$ and $\gamma  \otimes\delta$, then the set of 
$I^{(\alpha\beta\gamma\delta,\eta)}$ forms a complete family of $G$-invariants 
generating $U^{(\alpha\beta\gamma\delta)}$.

Moreover, the number $n^{(\alpha \beta \gamma \delta)}$ of $G$-invariants for the 
set $(\alpha \beta \gamma \delta)$ can be obtained by the character formula
\begin{eqnarray}
n^{(\alpha\beta\gamma\delta)}  &=& 
\frac{1}{|G|}\sum_{R \in G} \chi^\alpha(R)^* \chi^\beta(R)^* \chi^\gamma(R) \chi^\delta(R) ,
\label{eq:nbinvariants}
\end{eqnarray}
where the character of a representation $\eta$ is defined as 
$\chi^\eta(R)=\Tr \big(\Gamma^{(\eta)}(R)\big)$.

\section{Permutation-symmetrized invariants}
\label{sec:perminv}

The $G$-invariants of Eq.~\eqref{eq:definvariant} do not take into 
account the permutation symmetry as described in 
section~\ref{sec:symmetries_permutations}. These additional 
constraints considerably decrease the number of invariants and are 
discussed now.

\subsection{Real representations} \label{sec:perminv-real}

In this section, we assume that all the irreps are real and that they 
are represented by real matrices. Consequently, the Clebsch-Gordan 
coefficients can be chosen real, too. The suitable permutation invariance 
of the integrals were given in Eq.~\eqref{eq:permutations_real}.

\subsubsection{Symmetrization}

As for any group, we obtain $D_4$-symmetrized $G$-invariants by 
projecting $U_{abcd}^{(\alpha \beta \gamma \delta)}$ onto the 
fully-symmetric irrep $A_1$ of the permutation group $D_4$
\begin{eqnarray} \nonumber
\langle U^{(\alpha \beta \gamma \delta)}_{abcd} \rangle &=& 
\frac{1}{|D_4|} \sum_{p \in D_4} U_{p(abcd)}^{p(\alpha\beta\gamma\delta)} 
\label{symmetrizedI}\\
&=& \frac{1}{8} \big( U_{abcd}^{(\alpha\beta\gamma\delta)} + 
U_{cdab}^{(\gamma\delta\alpha\beta)} + U_{badc}^{(\beta\alpha\delta\gamma)} 
+ U_{dcba}^{(\delta\gamma\beta\alpha)} \nonumber \\ \nonumber
&& + U_{adcb}^{(\alpha\delta\gamma\beta)} + U_{cbad}^{(\gamma\beta\alpha\delta)} + 
U_{bcda}^{(\beta\gamma\delta\alpha)}  + U_{dabc}^{(\delta\alpha\beta\gamma)}  \big). \\ 
\end{eqnarray}
By definition, $\langle U^{(\alpha \beta \gamma \delta)}_{abcd} \rangle$ is 
invariant under the action of $D_4$. In fact, we know from 
Eq. \eqref{eq:permutations_real} that Coulomb integrals satisfy 
$\langle U^{(\alpha \beta \gamma \delta)}_{abcd} \rangle=
U^{(\alpha \beta \gamma \delta)}_{abcd}$, but as for the calculation
of $G$-invariants, we investigate the properties of $D_4$-symmetrized
Coulomb integrals by assuming that we start from non-symmetrized ones. 
As a consequence, the permutation-symmetrized $G$-invariants are given by
\begin{eqnarray}
\langle I^{(\alpha \beta \gamma \delta,\eta)} \rangle = \sum_{abcde} 
\frac{\langle U^{(\alpha \beta \gamma \delta)}_{abcd} \rangle 
(\alpha a \beta b|\eta e) (\gamma c \delta d|\eta e)}{\dim\eta},
\label{eq:rawGD4invariants}
\end{eqnarray}
which is a sum of 8 terms as in Eq.~\eqref{symmetrizedI}.
This is the expression of the $G \times D_4$ invariants in terms of the 
Coulomb integrals. We will write them in terms of the 
$G$-invariants, as detailed the next two subsections. 
The goal is to transform the Clebsch-Gordan coefficients of each 
sum so that they match the indices of the permuted integral
$U_{p(abcd)}^{p(\alpha\beta\gamma\delta)}$, in order to identify 
a $G$-invariant. Two cases must be distinguished: either the 
representations coupled by the Clebsch-Gordan coefficients are 
also coupled in the permuted integral (pair-conserving permutations), 
or they are reshuffled (pair-mixing permutations).

\paragraph{The first four terms: pair-conserving permutations.}

The terms of $\langle I^{(\alpha \beta \gamma \delta,\eta)} \rangle$
corresponding to the first two terms on the right hand side of 
Eq.~\eqref{symmetrizedI} are trivial
\begin{eqnarray*}
 \sum_{abcde} U_{abcd}^{(\alpha\beta\gamma\delta)}
(\alpha a \beta b|\eta e) (\gamma c \delta d|\eta e) &=& I^{(\alpha\beta\gamma\delta,\eta)}\dim\eta ,\\
\sum_{abcde} U_{cdab}^{(\gamma\delta\alpha\beta)}
 (\alpha a \beta b|\eta e) (\gamma c \delta d|\eta e) &=& I^{(\gamma\delta\alpha\beta,\eta)}\dim\eta .
\end{eqnarray*}
To deal with the next two terms, we notice that, as a consequence 
of Schur's lemma, the Clebsch-Gordan coefficients 
$(\alpha a \beta b|\eta e)$ and $(\beta b\alpha a |\eta e)$ differ by 
at most a phase depending on $\alpha$, $\beta$ and $\eta$ 
(but not on $a$, $b$ and $e$). We denote this phase by 
$\{\alpha\beta,\eta\}$~\cite{Piepho,Butler}: 
$(\alpha a \beta b|\eta e) = \{\alpha\beta,\eta\} (\beta b\alpha a |\eta e)$. 
Moreover $\{\alpha\beta,\eta\}=\pm1$ for real Clebsch-Gordan coefficients.

This enables us to write
\begin{eqnarray*}
&& \sum_{abcde}U_{badc}^{(\beta\alpha\delta\gamma)}
(\alpha a \beta b|\eta e)(\gamma c \delta d|\eta e) \\
&& = \{\alpha\beta,\eta\} \{\gamma\delta,\eta\}
\sum_{abcde}U_{badc}^{(\beta\alpha\delta\gamma)}
(\beta b\alpha a |\eta e) (\delta d\gamma c|\eta e) \\
&& = \{\alpha\beta,\eta\} \{\gamma\delta,\eta\} I^{(\beta\alpha\delta\gamma,\eta)} \dim\eta,
\end{eqnarray*}
and similarly
\begin{eqnarray*}
\sum_{abcde}U_{dcba}^{(\delta\gamma\beta\alpha)}
(\beta b\alpha a |\eta e) (\delta d\gamma c|\eta e)
&=&  \{\alpha\beta,\eta\} \{\gamma\delta,\eta\}
\\&&\times I^{(\delta\gamma\beta\alpha,\eta)}\dim\eta.
\end{eqnarray*}

\paragraph{The last four terms: pair-mixing permutations.}
\label{lastfoursect}
The last four terms are more cumbersome, because their Clebsch-Gordan 
coefficients couple the first and third, and the second and fourth 
representations of the integrals, thus breaking the ``bra'' and ``ket'' pairs.

In order to reshuffle these Clebsch-Gordan coefficients into new pairs, 
we need to use a recoupling formula. Precisely, we use the one proposed 
by Derome and Sharp (theorem 3 of~\cite{Derome-65}, and see notes~\cite{note2, note3}) 
for their general Clebsch-Gordan coefficients. For real representation matrices 
(and multiplicity-free point groups), the recoupling formula takes the form
\begin{eqnarray}
\sum_e
(\alpha a \beta b |\eta e)
(\gamma c\delta d|\eta e) &=& 
\dim \eta
\sum_{\phi f} (-1)^{\phi+\eta}  
\sixj{\alpha}{\beta}{\eta}{\gamma}{\delta }{\phi }
\nonumber\\&\times&
(\gamma c \beta b| \phi f)
(\alpha a \delta d | \phi f), 
\label{recoupling}
\end{eqnarray}
where the $6j$-symbols are defined by
\begin{eqnarray}
\sixj{\alpha}{\beta}{\eta}{\gamma}{\delta}{\phi} &= &
\frac{(-1)^{\alpha+\beta+\eta+\gamma+\delta+\phi}}{\dim\eta\dim\phi} \nonumber \\
&&\hspace{-33mm} \qquad \qquad \times  \sum_{abcdef}
(\gamma c\delta d|\eta e)
(\gamma c \beta b|\phi f)
(\alpha a \delta d| \phi f)
(\alpha a \beta b | \eta e).
\label{sixjdef}
\end{eqnarray}
In Eqs.~\eqref{recoupling} and \eqref{sixjdef}, the symbols 
$(-1)^\alpha$, $(-1)^\beta, \ldots$ are defined as follows. According to 
Derome and Sharp, we first define $1j$-symbols by 
$(\alpha)_{aa'}=(\alpha a 00|\alpha a')$, where $0$ is the fully-symmetric 
irrep. By Schur's lemma, it can be shown that $(\alpha)_{aa'}=\pm \delta_{aa'}$ 
and we denote the sign $\pm$ by $(-1)^\alpha$. In particular, if Clebsch-Gordan 
coefficients are calculated from Dirl's formula in Eq.~\eqref{eq:CGDirl}, then 
$(-1)^\alpha=1$ for every irrep $\alpha$. We also use the obvious notation 
$(-1)^{\phi+\eta} =(-1)^{\phi} (-1)^{\eta} $.

In the literature, other $6j$-symbols were defined which display interesting 
symmetry properties~\cite{Butler,Piepho}. However, these symmetries require 
to adjust the phases of the Clebsch-Gordan coefficients. Instead, we follow 
Derome's approach again and work with general Clebsch-Gordan 
coefficients~\cite{Derome-PhD, Derome-65, Derome-66}. We are now ready 
to apply the remaining permutations of $D_4$ to $I^{(\alpha\beta\gamma\delta,\eta)}$. 
For example
\begin{align*}
& \sum_{abcde} U_{cbad}^{(\gamma\beta\alpha\delta)}
(\alpha a \beta b|\eta e)(\gamma c \delta d|\eta e)
\\
&= \dim \eta
\sum_{\phi} (-1)^{\phi+\eta}  
 \sixj{\alpha}{\beta}{\eta}{\gamma}{\delta }{\phi } \\
& \quad \times \sum_{abcdf}  U_{cbad}^{(\gamma\beta\alpha\delta)}
(\gamma c \beta b| \phi f)
(\alpha a \delta d | \phi f) \\
&= \dim \eta \sum_{\phi} (-1)^{\phi+\eta}  \dim\phi
\sixj{\alpha}{\beta}{\eta}{\gamma}{\delta }{\phi }
I^{(\gamma\beta\alpha\delta,\phi)}.
\end{align*}

\paragraph{\texorpdfstring{$D_4$}{D4}-symmetrized invariant.}
By treating the three remaining terms in the same manner, the 
$D_4$-symmetrization of $I^{(\alpha\beta\gamma\delta,\eta)}$ yields
\begin{eqnarray}
\langle I^{(\alpha\beta\gamma\delta,\eta)}\rangle &=& 
\frac{1}{8}\Big(I_{pc}
 +
 \sum_{\phi} (-1)^{\eta+\phi}\dim\phi 
\sixj{\alpha}{\beta}{\eta}{\gamma}{\delta}{\phi}
I_{npc} \Big),
\nonumber\\\label{symIdetail}
\end{eqnarray}
where the pair-conserving terms $I_{pc}$
and the pair-non-conserving terms $I_{npc}$ are
\begin{eqnarray*}
I_{pc} &=&
I^{(\alpha\beta\gamma\delta,\eta)}+
I^{(\gamma\delta\alpha\beta,\eta)}
\\&&
 + \{\alpha\beta,\eta\} \{\gamma\delta,\eta\} \big(
      I^{(\beta\alpha\delta\gamma,\eta)} 
+ I^{(\delta\gamma\beta\alpha,\eta)}\big),
\end{eqnarray*}
\begin{eqnarray*}
I_{npc} &=&
I^{(\alpha\delta\gamma\beta,\phi)}  + I^{(\gamma\beta\alpha\delta,\phi)}
\\&&
+ \{\alpha\delta,\phi \} \{\beta\gamma,\phi \} 
 \big( I^{(\beta\gamma\delta\alpha,\phi)} + I^{(\delta\alpha\beta\gamma,\phi)}
\big).
\end{eqnarray*}

\subsubsection{Enumeration of \texorpdfstring{$G \times D_4$}{GxD4}-invariants}
\label{enumeration-sect}

A permutation $p$ of $D_4$ transforms a quadruple of irreps 
$\sigma=(\alpha,\beta,\gamma,\delta)$ into another quadruple
$p(\sigma)$. 
Let $S$ be the set of quadruples obtained from 
$\sigma$ by the action of $D_4$ (this is called the \textit{orbit} of $\sigma$).
The number of elements of $S$ depends on the irreps in $\sigma$. 
For instance, if $\sigma=(\alpha,\alpha,\alpha,\alpha)$, then $S$
has only one element $S=\{\sigma\}$. If all irreps are different,
then $S$ has 8 elements. We shall see that there are five different
types of $S$.
 
As noticed in Ref.~\cite{Bunemann-17}, the number of independent 
permutation-symmetrized $G \times D_4$-invariants for a given set $S$ 
can again be obtained from the character formula
\begin{equation}
n^S = \frac{1}{|G \times D_4|} \sum_{(R,p) \in G \times D_4} \chi^S(R,p) ,
\label{eq:nbinvariants_GGP}
\end{equation}
where the character $\chi^S(R,p)$ of the element $(R,p)$ in 
$G \times D_4$ in the orbit $S$ is defined by
\begin{eqnarray*}
&& \chi^{S}(R,p) =
 \sum_{(\alpha\beta\gamma\delta)\in S}
\delta_{(\alpha\beta\gamma\delta),p(\alpha\beta\gamma\delta)} \\
&& \quad \times \sum_{abcd}
\Gamma_{a'a}^{(\alpha)}(R) 
\Gamma_{b'b}^{(\beta)}(R) 
\Gamma_{c'c}^{(\gamma)}(R) 
\Gamma_{d'd}^{(\delta)}(R) \Big|_{(a'b'c'd')=p^{-1}(abcd)}.
\end{eqnarray*}

We give the formula for $\sum_p \chi^{S}(R,p)$ in terms of the 
characters $\chi^\alpha(R)$ of $G$ for every possible set $S$.
\begin{itemize}
\item $S_1=\{(\alpha,\alpha,\alpha,\alpha)\}$
\begin{eqnarray*}
\sum_p\chi^{S_1}(R,p) &=& 
\chi^{\alpha}(R)^4+3\chi^{\alpha}(R^2)^2 \\
&& + 2\chi^{\alpha}(R^2)\chi^{\alpha}(R)^2 +2\chi^{\alpha}(R^4) ;
\end{eqnarray*}
\item $S_2=\{(\alpha,\beta,\alpha,\beta),(\beta,\alpha,\beta,\alpha)\}$ with $\beta\not=\alpha$
\begin{eqnarray*}
\sum_p\chi^{S_2}(R,p) &=& 
2\big(\chi^{\alpha}(R)^2 +\chi^{\alpha}(R^2)\big) \\
&& \quad \times \big(\chi^{\beta}(R)^2 +\chi^{\beta}(R^2)\big) ;
\end{eqnarray*}
\item $S_3=\{(\alpha,\alpha,\beta,\beta), (\alpha,\beta,\beta,\alpha), (\beta,\beta,\alpha,\alpha),
 (\beta,\alpha,\alpha,\beta)\}$ with $\beta\not=\alpha$
\begin{eqnarray*}
\sum_p\chi^{S_3}(R,p) &=& 
4\chi^{\alpha}(R)^2 \chi^{\beta}(R)^2+4\chi^{\alpha}(R^2)\chi^{\beta}(R^2) ;  \\
\end{eqnarray*}
\item $S_4=\{(\alpha,\beta,\alpha,\gamma), (\alpha,\gamma,\alpha,\beta), 
(\beta,\alpha,\gamma,\alpha), (\gamma,\alpha,\beta,\alpha)\}$ where 
$\beta\not=\alpha$ and $\beta\not=\gamma$, but $\gamma=\alpha$ is allowed
\begin{eqnarray*}
\sum_p\chi^{S_4}(R,p) &=&
4\big(\chi^{\alpha}(R)^2 +
\chi^{\alpha}(R^2)\big)\chi^{\beta}(R)\chi^{\gamma}(R) ; \nonumber \\
\end{eqnarray*}
\item For all the other cases
\begin{eqnarray*}
\sum_p\chi^{S_5}(R,p) &=& 8\chi^{\alpha}(R) \chi^{\beta}(R)\chi^{\gamma}(R)\chi^{\delta}(R).
\end{eqnarray*}
\end{itemize}

\subsubsection{Independent components of Coulomb integrals}

For convenience, we write Eq.~\eqref{symIdetail} in matrix form: $\langle I^p \rangle=M_{pq} I^q$,
where $p$ and $q$ are compound indexes $\sigma,\eta$
and were $M^2=M$ is a projection matrix. 
As such, it is diagonalizable, but
since $M$ is not an orthogonal projection, its eigenspaces
are not orthogonal. To solve this problem, we define
the matrix $Q_{(\sigma,\eta)(\tau,\phi)}=\delta_{\sigma\tau}\delta_{\eta\phi}
\sqrt{\dim\eta}$ and the modified projection
matrix $N=Q M Q^{-1}$, which 
is symmetric
(i.e. $N^T=N$)
because of the 
symmetries of $6j$ symbols~\cite{Derome-65}. The eigenvalues of $N$ are 0 and 1
and its orthonormal eigenvectors are denoted by $\bfv^p$. 

In general, when a matrix $N$ is diagonalizable
with eigenvalues $\lambda_p$ and eigenvectors $\bfv^p$,
we can define the matrix $B_{qp}=v^p_q$, where $v^p_q$ is
the $q$th component of $\bfv^p$ and $N$ is recovered
by $N_{pr}=\sum_q B_{pq} \lambda_q (B^{-1})_{qr}$. 
The eigenvalue $\lambda_q=0$ does obviously not contribute and 
we are left with $N_{pr}=\sum_{q,\lambda_q=1} B_{pq} (B^{-1})_{qr}$.
We can now define the \textit{independent components} $u^q$ 
of symmetrized  $G$-invariants to be
\begin{eqnarray}
u^q &=& \sum_r (B^{-1}Q)_{qr} I^r,
\label{defuq}
\end{eqnarray}
where $q$ is such as $\lambda_q=1$.
These independent components are the minimal information
required to compute all Coulomb integrals.
Indeed
\begin{eqnarray}
 \sum_{q,\lambda_q=1} (Q^{-1} B)_{pq} u^q
&=&
  \sum_{r}  M_{pr} I^r
=\langle I^p\rangle.
\label{moyIu}
\end{eqnarray}

\subsubsection{The norm of Coulomb integrals}
\label{distance-sect}

In order to analyze the screening of electron-electron interaction in 
the solid-state, evaluate the effect of an external parameter 
such as pressure on Coulomb integrals, or simply 
fit  Coulomb integrals in a crystal with a spherical model, we 
need to evaluate the distance between Coulomb integrals.
Since Coulomb integrals are matrix elements of an operator,
the natural distance is given by the 
Hilbert-Schmidt (or Frobenius) norm 
 defined by
\begin{eqnarray*}
||U^{(\alpha\beta\gamma\delta)}||^2 &=& 
\sum_{abcd} |U^{(\alpha\beta\gamma\delta)}_{abcd}|^2.
\end{eqnarray*}
This norm is natural because it is invariant under unitary 
transformations and it is perfectly suited to least square
minimization. 
The mixing of irreps due to permutations
leads us to consider also the distance between Coulomb integrals
for a given set $S$ of quadruples $\sigma=(\alpha\beta\gamma\delta)$
\begin{eqnarray*}
||U^S||^2 &=& 
\sum_{\sigma \in S} ||U^\sigma||^2.
\end{eqnarray*}

Since we express Coulomb integrals in terms of $G$-invariants,
we need to define a distance between $G$-invariants which is 
compatible with the distance between Coulomb integrals. 
By using Eq.~\eqref{eq:Ufrominvariant} and the orthogonality
of Clebsch-Gordan coefficients we obtain
\begin{eqnarray}
||U^S||^2 &=& \sum_{\sigma\in S}\sum_\eta
\dim\eta |I^{(\sigma,\eta)}|^2.
\label{distUSI}
\end{eqnarray}
We are now ready to compute the norm of Coulomb integrals
in terms of the independent components. 
Since matrix $Q^{-1} $ in Eq.~\eqref{moyIu} removes the coefficient $\dim\eta$ 
in Eq.~\eqref{distUSI} and the columns of $B$ are orthonormal, we find
\begin{eqnarray}
||U^S||^2 &=& \sum_{q,\lambda_q=1}
|u^q|^2,
\label{distUSu}
\end{eqnarray}
where, in Eq.~\eqref{distUSI}, we used $I^p=\langle I^p\rangle$,
which is a consequence of definition~\eqref{eq:rawGD4invariants} 
of $\langle I^p\rangle$
when Coulomb integrals satisfy the $D_4$-permutation
symmetry (i.e. $ U^{(\alpha \beta \gamma \delta)}_{abcd}=
\langle U^{(\alpha \beta \gamma \delta)}_{abcd} \rangle$).

In summary, it is possible to evaluate the total distance between
two sets of Coulomb integrals from the distance between the independent 
components $u^q$. An example of this calculation is given in 
section~\ref{perO3-sect}.

\subsubsection{Minimizing Coulomb integral calculations}
\label{minimizing-section}

In this section, we determine the minimal number of Coulomb integrals
$U^{(\alpha\beta\gamma\delta)}_{abcd}$ that we have to calculate
to be able to determine all Coulomb integrals. This is obviously useful to 
mimize the computational cost of electronic structure calculations.
We recall Eq.~\eqref{eq:Ufrominvariant}
\begin{eqnarray*}
U_{abcd}^{(\alpha\beta\gamma\delta)}  &=& 
\sum_{\eta e} (\alpha a \beta b|\eta e) (\gamma c \delta d|\eta e)
I^{(\alpha\beta\gamma\delta,\eta)},
\end{eqnarray*}
which can be used to calculate 
 $n=\dim\alpha\dim\beta\dim\gamma\dim\delta$
Coulomb integrals in terms of $n^{(\alpha \beta \gamma \delta)}$ 
$G$-invariants (see Eq.~\eqref{eq:nbinvariants}).
If $S$ denotes the type of set described in section~\ref{enumeration-section} to which
$(\alpha\beta\gamma\delta)$ belongs, symmetrization
mixes now $|S|n$ Coulomb integrals, where $|S|$ is the number
of elements of $S$, and these $|S|n$ Coulomb integrals
can be determined in terms of $n^S$ permutation-symmetrized
$G$-invariants, as explained in section~\ref{enumeration-section}.
More precisely, we have shown in the previous section that
there are $n^S$ quantities $u^q$ such that
$\langle I^p\rangle = \sum_{q} (Q^{-1}B)_{pq}u^q$.
Combining these two relations, we see that there
is an $(|S|n)\times n^S$ matrix $A^{\alpha\beta\gamma\delta,q}_{abcd}$, 
where $(\alpha\beta\gamma\delta)$ is in $S$, such that
\begin{eqnarray*}
U_{abcd}^{(\alpha\beta\gamma\delta)}  &=& 
\sum_q A^{\alpha\beta\gamma\delta,q}_{abcd} u^q.
\end{eqnarray*}

The rank $r$ of $A$ is the length of the longest list of independent columns
of $A$~\cite{Horn2}. The case $r < n^S$ is possible, 
but it means that the Coulomb
integrals are linear combinations of a number of parameters smaller
than $n^S$. This can happen when the true symmetry group is
larger than the one considered in the calculation. However the
most common case is $r=n^S$ and we only consider this case.
By definition of the rank of $A$, we can choose $n^S$ independent columns of
$A$. Each column corresponds to a specific
Coulomb integral $U_{a_pb_pc_pd_p}^{(\alpha_p\beta_p\gamma_p\delta_p)}$.
We take such a set of independent columns to build 
an $n^S\times n^S$ matrix $P$ relating the 
$U_{a_pb_pc_pd_p}^{(\alpha_p\beta_p\gamma_p\delta_p)}$
to $u^q$. The matrix $P$ is invertible since the columns are independent.
Therefore, $P^{-1}$ allows us to express 
the $n^S$ independent components $u^q$ in terms of the $n^S$
selected Coulomb integrals. 
Since the independent components generate all Coulomb integrals, we can
compute all Coulomb integrals from $n^S$ of them.

An example of this selection of a minimum of Coulomb integrals
 is given in section~\ref{perO3-sect}.

\subsection{Non-real representations}

In this section, the representation matrices, the orbital wavefunctions 
and the Clebsch-Gordan coefficients are assumed to be complex. The 
suitable permutation invariance of the integrals were given in 
Eq.~\eqref{eq:permutations_nonreal}.

\subsubsection{Symmetrization}

It is clear that 
\begin{eqnarray*}
\langle U_{abcd}^{(\alpha\beta\gamma\delta)}\rangle
&=& \frac14\Big( U_{abcd}^{(\alpha\beta\gamma\delta)} 
+U_{badc}^{(\beta\alpha\delta\gamma)}  \\
&& \qquad +(U_{cdab}^{(\gamma\delta\alpha\beta)})^*
+(U_{dcba}^{(\delta\gamma\beta\alpha)})^*\Big),
\end{eqnarray*}
is invariant under the operations of $m'm2'$.
Therefore, the relation
between symmetrized
and non-symmetrized components for non-real representations is
\begin{eqnarray*}
4\langle I^{(\alpha\beta\gamma\delta,\eta)}\rangle &=&
 \sum_{abcde}U_{abcd}^{(\alpha\beta\gamma\delta)}
(\alpha a \beta b|\eta e)^*(\gamma c \delta d|\eta e)
\\&&
+ \sum_{abcde}U_{badc}^{(\beta\alpha\delta\gamma)}
(\alpha a \beta b|\eta e)^*(\gamma c \delta d|\eta e)
\\&&
+ \sum_{abcde}(U_{cdab}^{(\gamma\delta\alpha\beta)})^*
(\alpha a \beta b|\eta e)^*(\gamma c \delta d|\eta e)
\\&&
+ \sum_{abcde}(U_{dcba}^{(\delta\gamma\beta\alpha)})^*
(\alpha a \beta b|\eta e)^*(\gamma c \delta d|\eta e)
 \\
&=& I^{(\alpha\beta\gamma\delta,\eta)}
+\{\alpha\beta\eta\}\{\gamma\delta\eta\}
I^{(\beta\alpha\delta\gamma,\eta)}
\\&&
+(I^{(\gamma\delta\alpha\beta,\eta)})^*
+\{\alpha\beta\eta\}\{\gamma\delta\eta\}
(I^{(\delta\gamma\beta\alpha,\eta)})^*.
\end{eqnarray*}
By similarly calculating $\langle I^{(\beta\alpha\delta\gamma,\eta)}\rangle$,
$\langle I^{(\gamma\delta\alpha\beta,\eta)} \rangle$
and $\langle I^{(\delta\gamma\beta\alpha,\eta)}\rangle$, we obtain
 the following relations between symmetrized components
\begin{eqnarray*}
\langle  I^{(\beta\alpha\delta\gamma,\eta)}\rangle
&=& \{\alpha\beta\eta\}^*\{\gamma\delta\eta\}
\langle I^{(\alpha\beta\gamma\delta,\eta)}\rangle,\\
\langle  I^{(\gamma\delta\alpha\beta,\eta)}\rangle &=&
\langle I^{(\alpha\beta\gamma\delta,\eta)}\rangle^*,\\
\langle  I^{(\delta\gamma\beta\alpha,\eta)}\rangle &=&
\{\alpha\beta\eta\}^*\{\gamma\delta\eta\} 
\langle I^{(\alpha\beta\gamma\delta,\eta)}\rangle^*.
 \end{eqnarray*}

\subsubsection{Enumeration of symmetrized \texorpdfstring{$G$}{G}-components}
\label{enumeration-section}

Irreducible corepresentations are not as familiar as irreps but
there is also a character formula for counting the number of times 
a given irreducible corepresentation appears in a
general corepresentation~\cite{Newmarch-83}.
In Newmarch's language, we consider the fully symmetric 
irreducible corepresentation, which is of type $(a)$, 
corresponding to an intertwining number $I=1$. 

A particularity of the character theory of corepresentations is 
that it takes into account
only unitary operations (in our case the unit permutation
$(\alpha\beta\gamma\delta)$ and the
permutation $(\beta\alpha\delta\gamma)$). 
The character of the representation corresponding to permutation $p$ is
\begin{eqnarray*}
\chi^{S}(R,p) &=& 
\sum_{(\alpha\beta\gamma\delta)\in S}
\delta_{(\alpha\beta\gamma\delta),p(\alpha\beta\gamma\delta)} \\
&& \! \! \! \! \! \! \! \! \! \! \! \! \! \! \! \! \! \!  \! \! \! \! \!  \! \! \! \! \!  \times
\sum_{abcd}
\Gamma_{a'a}^{(\alpha)}(R)^*
\Gamma_{b'b}^{(\beta)}(R)^*
\Gamma_{c'c}^{(\gamma)}(R) 
\Gamma_{d'd}^{(\delta)}(R)\Big|_{(a'b'c'd')=p^{-1}(abcd)},
\end{eqnarray*}
and the number of symmetrized $G$-components is
\begin{eqnarray*}
n^S &=& \frac{1}{2|G|}{\sum_p}' \sum_{R} \chi^{S}(R,p),
\end{eqnarray*}
where $\sum'$ runs only over the two permutations corresponding to
unitary  operations.
We have only two cases to consider
\begin{itemize}
\item $S=\{(\alpha,\alpha,\beta,\beta)\}$, where $\alpha$ and $\beta$ can be equal
\begin{eqnarray}
n^S &=& 
 \frac{1}{2|G|}\sum_{R} \Big(
\big(\chi^{\alpha}(R)^*\big)^2 \chi^{\beta}(R)^2+
\chi^{\alpha}(R^2)^*\chi^{\beta}(R^2)\Big) ; \nonumber \\
\end{eqnarray}
\item For all other cases
\begin{eqnarray}
n^S &=& 
 \frac{1}{|G|}\sum_{R} 
\chi^{\alpha}(R)^*\chi^{\beta}(R)^*
\chi^{\gamma}(R)\chi^{\delta}(R).
\end{eqnarray}
\end{itemize}

\subsubsection{Minimizing Coulomb integral calculations}

Exactly as in the case of real representation matrices treated in 
section~\ref{minimizing-section}, Coulomb integrals
can be calculated from a minimum number $n^S$ of them.

\section{Subduction}
\label{sec:subd}

In this section, we consider the point group
$G$, which
is a subgroup of a larger group $\calG$
(symmetry breaking) in order
to compare the invariants of both groups. To do so, we 
will give the expression of the $G$-invariants on the basis of the $\calG$-invariants.

A typical application consists in taking the continuous rotation group 
$SO(3)$ as the larger group. Therefore, we first show that the $SO(3)$ 
invariants are related to the well-known Slater integrals, which parametrize 
Coulomb interaction in spherical symmetry. 

All point symmetry groups are subgroups of 
$O(3)$ rather than $SO(3)$, but since $O(3)$ is the direct product
of $SO(3)$ and $C_i\{1,\mathcal{I}\}$, where $\mathcal{I}$ is the inversion, the
irreps of $O(3)$ are direct products of irreps of $SO(3)$ and
of $C_i$. To simplify notations, we first concentrate on $SO(3)$
and use the results to describe subduction from $O(3)$.

\subsection{\texorpdfstring{$SO(3)$}{SO(3)}-invariants\label{sec:SO3invariants}}

\subsubsection{\texorpdfstring{$SO(3)$}{SO(3)}-invariants for spherically-symmetric potentials}

The theory presented in the previous section does not directly apply to
$SO(3)$ because, although irreps $\ell$ are real (in the sense of Frobenius-Schur),
they are usually represented by Wigner matrices
$D^{\ell}(R)$ which can be complex. The usual basis of 
spherical harmonics $Y_\ell^m$ is also generally
not real. This, however, has only a benign effect and we only indicate the results.
We follow the notation used in Cowan's book~\cite{Cowan},
where Coulomb integrals are given (for a Coulomb potential) by
\begin{eqnarray*}
U^{(\ell_1\ell_2\ell_3\ell_4)}_{m_1 m_2 m_3 m_4} &=&
\langle\ell_1 m_1\ell_2 m_2|\frac{2}{r_{ij}}|\ell_3 m_3\ell_4 m_4\rangle
\\&=&
\sum_k R^k(\ell_1\ell_2,\ell_3\ell_4) \delta_{m_3-m_1,m_2-m_4}
(-1)^{m_2+m_3} 
\\&&
\times \sqrt{(2\ell_1+1)(2\ell_2+1)(2\ell_3+1)(2\ell_4+1)}
\\&&
\times \threej{\ell_1}{k}{\ell_3}{0}{0}{0}
\threej{\ell_2}{k}{\ell_4}{0}{0}{0}
\\&&
\times \threej{\ell_1}{k}{\ell_3}{-m_1}{m_1{-}m_3}{m_3}
\threej{\ell_2}{k}{\ell_4}{-m_2}{m_2{-}m_4}{m_4},
\end{eqnarray*}
where~\cite{Cowan}
\begin{eqnarray*}
R^k(\ell_1\ell_2,\ell_3\ell_4) &=&
\int_0^\infty r_1^2 dr_1 
\int_0^\infty r_2^2 dr_2 \frac{2r_<^k}{r_>^{k+1}}
\nonumber\\
&& \qquad  \times 
 R_{\ell_1}(r_1) R_{\ell_2}(r_2) R_{\ell_3}(r_1) R_{\ell_4}(r_2), 
\end{eqnarray*}
are radial integrals and we assumed real radial wavefunctions
$R_\ell$.
The $3j$-symbols involving a row of zeros can only be non zero if 
$\ell_1+\ell_3+k$ and $\ell_2+\ell_4+k$ are even.
As a consequence, $\ell_1+\ell_3+\ell_2+\ell_4$ is even
and $(-1)^{\ell_1+\ell_3+\ell_2+\ell_4}=1$.

Our $SO(3)$-invariants $I^{(\ell_1 \ell_2 \ell_3 \ell_4,\ell)}$
were calculated by Cowan (Eq.~(10.17) and (10.20), ~\cite{Cowan})
\begin{eqnarray*}
I^{(\ell_1 \ell_2 \ell_3 \ell_4,\ell)} &=&
 \langle(\ell_1\otimes \ell_2)^\ell_m|\frac{2}{r_{ij}}|(\ell_3\otimes \ell_4)^\ell_m\rangle 
\\&=&
 (-1)^{\ell_1-\ell_3+\ell} \\
&&\times \sqrt{(2\ell_1+1)(2\ell_2+1)(2\ell_3+1)(2\ell_4+1)} \\
&&\times \sum_k
\threej{\ell_1}{k}{\ell_3}{0}{0}{0}
\threej{\ell_2}{k}{\ell_4}{0}{0}{0}
\sixj{\ell_1}{\ell_2}{\ell}{\ell_4}{\ell_3}{k} \\
&&\times R^k(\ell_1\ell_2,\ell_3\ell_4),
\end{eqnarray*}
where the right hand side is known to be independent of $m$.
Cowan's result relates to the present work through
\begin{eqnarray*}
|(\ell_1\otimes \ell_2)^\ell_m\rangle=\sum_{m_1m_2}
(\ell_1m_1\ell_2m_2|\ell m)\varphi^{\ell_1}_{m_1}(\bfr_i)
\varphi^{\ell_2}_{m_2}(\bfr_j). 
\end{eqnarray*}
Moreover, since the right hand side of the
equation for $I^{(\ell_1 \ell_2 \ell_3 \ell_4,\ell)}$ 
is independent of $m$, it can be replaced by
its average over $m$. Thus, we obtain
\begin{eqnarray*}
I^{(\ell_1 \ell_2 \ell_3 \ell_4,\ell)} &=&
\frac{1}{2\ell+1}\sum_{m_1m_2 m_3 m_4 m}
(\ell_1 m_1 \ell_2 m_2|\ell m) \\
&&\qquad \qquad  \times (\ell_3 m_3 \ell_4 m_4|\ell m)
\,U^{(\ell_1\ell_2\ell_3\ell_4)}_{m_1 m_2 m_3 m_4},
\end{eqnarray*}
which is indeed a 
$SO(3)$-invariant in the sense of Eq.~\eqref{eq:definvariant}.

In the case of $d$~electrons for which $\ell_1=\ell_2=\ell_3=\ell_4=2$,
we write $F^k=R^k(22,22)$ for the standard $d$-shell
Slater integrals and we obtain
\begin{eqnarray*}
I^{(2^4,0)} &=& F^0 + \frac27 F^2 + \frac27 F^4,\\
I^{(2^4,1)} &=& F^0 + \frac17 F^2 - \frac{4}{21} F^4,\\ 
I^{(2^4,2)} &=& F^0 - \frac{3}{49}  F^2 +\frac{4}{49} F^4,\\
I^{(2^4,3)} &=& F^0 - \frac{8}{49}  F^2 -\frac{1}{49} F^4,\\ 
I^{(2^4,4)} &=& F^0 + \frac{4}{49}  F^2 +\frac{1}{441} F^4.
\end{eqnarray*}

The expressions of this section were obtained by explicitly assuming that
the interaction potential is the Coulomb potential. In the following, we 
only assume that the potential is spherically symmetric (so that the invariants remain 
$SO(3)$-invariants) and symmetric under the exchange of particles (so that the $D_4$
permutation symmetry is still valid.) This enables us to consider more general effective 
potentials.

\subsubsection{Permutation-symmetrized \texorpdfstring{$SO(3)$}{SO(3)}-invariants}
\label{perO3-sect}

As in the case of the general group $G$, we can build 
permutation-symmetrized $SO(3)$-invariants. The $D_4$ 
permutation symmetry can only be applied if the complex spherical harmonics are 
transformed into real (or cubic or tesseral) harmonics, but the corresponding 
Clebsch-Gordan coefficients are not the usual ones. We prefer to work with spherical 
harmonics, but the action of $D_4$ permutations is now different
\begin{eqnarray*}
U^{(\ell_1\ell_2\ell_3\ell_4)}_{m_1 m_2 m_3 m_4} &=&
U^{(\ell_2\ell_1\ell_4\ell_3)}_{m_2 m_1 m_4 m_3} 
=(-1)^{m_1+m_3} U^{(\ell_2\ell_3\ell_4\ell_1)}_{ m_2 -m_3 m_4 -m_1 }\\
&=&
(-1)^{m_1+m_2+m_3+m_4}U^{(\ell_3\ell_4\ell_1\ell_2)}_{ -m_3 -m_4-m_1 -m_2}
\\&=&
(-1)^{m_1+m_2+m_3+m_4 } U^{(\ell_4\ell_3\ell_2\ell_1)}_{ -m_4 -m_3 -m_2 -m_1} 
\\&=&
(-1)^{m_1+m_3} U^{(\ell_3\ell_2\ell_1\ell_4)}_{-m_3 m_2 -m_1 m_4}
\\&=&
(-1)^{m_2+m_4} U^{(\ell_4\ell_1\ell_2\ell_3)}_{-m_4 m_1 -m_2 m_3 } 
\\&=&
(-1)^{m_2+m_4} U^{(\ell_1\ell_4\ell_3\ell_2)}_{ m_1 -m_4 m_3 -m_2}.
\end{eqnarray*}
Still, the result $\langle I^{(\ell_1 \ell_2 \ell_3 \ell_4,\ell)}\rangle$ 
of permutation symmetrization
has the same form as the one for real representation matrices given by
Eq.~\eqref{symIdetail}, if we substitute $\alpha=\ell_1$, $\beta=\ell_2$,
$\gamma=\ell_3$, $\delta=\ell_4$, $\eta=\ell$, $\phi=\ell'$,
all permutation factors $\{\ell_i\ell_j,\ell_k\}=(-1)^{\ell_i+\ell_j-\ell_k}$ 
and $(-1)^{\eta+\phi}$ is replaced by $(-1)^{\ell_1+\ell_3}$.
Note that, because of the last substitution, the formula
for $SO(3)$ is not a special case of the general formula given 
in Eq.~\eqref{symIdetail} because,
for all irreps $\alpha(=\ell)$ of $SO(3)$,
 $(-1)^\alpha=1$
in the sense of the definition given in section~\ref{lastfoursect}. 
The additional factor $(-1)^{\ell_1+\ell_3}$ comes from
the fact that the action of $D_4$ involves signs
due to the complex nature of the representation matrices.

In particular, if $\ell_1=\ell_2=\ell_3=\ell_4$, then
\begin{eqnarray*}
\langle I^{(\ell_1^4,\ell)}\rangle
&=& 
\frac12 I^{(\ell_1^4,\ell)} +
\frac12
\sum_{\ell'} (2\ell'+1)
\sixj{\ell_1}{\ell_1}{\ell'}{\ell_1}{\ell_1}{\ell}  I^{(\ell_1^4,\ell')} .
\end{eqnarray*}

For $\ell_1=\ell_2=\ell_3=\ell_4=2$, this formula gives us 
\begin{eqnarray*}
\langle I^{(2^4,\ell)}\rangle
&=& 
\sum_{\ell'=0}^4  M_{\ell\ell'} I^{(2^4,\ell')},
\end{eqnarray*}
where 
\begin{eqnarray}
M
&=& \frac{1}{140}
\begin{pmatrix}
84 & -42 & 70 & -98 & 126 \\
-14 & 105 & -35 & 0 & 84 \\
14 & -21 & 55 & 56 & 36 \\
-14 & 0   & 40 & 105 &  9 \\
14 & 28 & 20 & 7 & 71
\end{pmatrix}.
\label{MO3}
\end{eqnarray}
As expected $M$ is a (non-orthogonal) projector (i.e. $M^2=M$) 
with eigenvalues $(1,1,1,0,0)$. Therefore, there are three independent 
components that can be related to the three Slater integrals. The reader 
can check that $\langle I^{(2^4,\ell)}\rangle= I^{(2^4,\ell)}$ when
$I^{(2^4,\ell)}$ is expressed in terms of Slater integrals
$F^k$ as in the end of the previous section.

To illustrate the construction described in section~\ref{distance-sect},
we consider the matrix $Q_{\ell\ell'}=\delta_{\ell\ell'}\sqrt{2\ell+1}$
and we choose three orthonormal eigenvectors $\bfv^{1,p}$ of
$N=QMQ^{-1}$ for eigenvalue 1 and two eigenvectors
$\bfv^{0,p}$ for eigenvalue 0 
(that we did not orthonormalize
to simplify its form)
to build the matrix
\begin{eqnarray*}
B &=& 
\left(
\begin{array}{ccccc}
 \frac{1}{5} & \frac{1}{\sqrt{5}} & \frac{3}{5} & -\frac{1}{2} & \frac{5}{2 \sqrt{7}} \\
 \frac{\sqrt{3}}{5} & \frac{\sqrt{\frac{3}{5}}}{2} & -\frac{2 \sqrt{3}}{5} & -\frac{5}{4 \sqrt{3}} &
   -\frac{\sqrt{\frac{3}{7}}}{4} \\
 \frac{1}{\sqrt{5}} & -\frac{3}{14} & \frac{6}{7 \sqrt{5}} & -\frac{\sqrt{5}}{4} & -\frac{\sqrt{35}}{4}
   \\
 \frac{\sqrt{7}}{5} & -\frac{4}{\sqrt{35}} & -\frac{3}{10 \sqrt{7}} & 0 & 1 \\
 \frac{3}{5} & \frac{6}{7 \sqrt{5}} & \frac{1}{70} & 1 & 0 \\
\end{array}
\right),
\end{eqnarray*}
giving us the three independent components
\begin{eqnarray}
u^1 &=& 5 F^0,\quad
u^2 = \frac{2\sqrt{5}}{7} F^2,\quad
u^3 = \frac{10}{21}  F^4.
\label{basisO3}
\end{eqnarray}
Eigenvectors $\bfv^{1,p}$ were chosen to
get this simple relation between independent components and Slater integrals.
By using Eq.~\eqref{distUSu},
we can now easily calculate the norm of the set
of 625 Coulomb integrals for $S=\{2,2,2,2\}$  in terms of Slater integrals
\begin{eqnarray*}
||U^S||^2 &=& 
\sum_{m_1 m_2 m_3 m_4}
|U^{(2^4)}_{m_1 m_2 m_3 m_4}|^2
\\&=&
25
(F^0)^2 + 
\frac{20}{49} (F^2)^2
+
\frac{100}{441} (F^4)^2.
\end{eqnarray*}

Incidentally, we recover the fact, often used in 
practice~\cite{Vaugier-12,Panda-17,vanRoekeghem-16} that spherically 
symmetric potentials can be described by the usual three Slater integrals 
$F^0$, $F^2$ and $F^4$ for $d$ orbitals. Indeed, Slater integrals were 
originally derived under the assumption that the interaction potential is of 
Coulomb type but the conclusion that there are only 3 symmetrized $SO(3)$ 
invariants are obtained here for more general potentials. We just assumed that 
the potential is real, spherically symmetric and invariant under the exchange 
of $\bfx$ and $\bfy$, so that the $D_4$ symmetry holds. Any real potential of 
the form $V(|\bfx|,|\bfy|,\bfx\cdot\bfy)=V(|\bfy|,|\bfx|,\bfx\cdot\bfy)$ satisfies 
these assumptions. 

We can also take advantage of this example to show how
the method described in section~\ref{minimizing-section}
minimizes the calculation of Coulomb integrals. 
Cowan's formula gives 625 Coulomb integrals 
$U^{(2^4)}_{m_1m_2m_3m_4}$ as a linear combination
of three Slater integrals, which are simply related to
our independent components in Eq.~\eqref{basisO3}. 
Therefore, the matrix $A$ relating Coulomb integrals
to independent components has 3 lines and 625 columns.
From this matrix we extract three columns
corresponding to $U^{(2^4)}_{0000}$,
$U^{(2^4)}_{001-1}$ and 
$U^{(2^4)}_{002-2}$ to build the $3\times 3$
matrix
\begin{eqnarray*}
P &=& \begin{pmatrix}
\frac{1}{5} & \frac{2}{7\sqrt{5}} & \frac{6}{35}\\
0 & -\frac{1}{14\sqrt{5}} & -\frac{1}{7}\\
0 & \frac{2}{7\sqrt{5}} & \frac{1}{14}
\end{pmatrix}.
\end{eqnarray*}
Since $\det P \not=0$, $P$ is invertible 
and we can compute the independent components
 $u^1$, $u^2$ and $u^3$ from
the Coulomb integrals
$U^{(2^4)}_{0000}$,
$U^{(2^4)}_{001-1}$ and 
$U^{(2^4)}_{002-2}$. Once we know the independent components, we
can calculate all 625 Coulomb integrals.

\subsubsection{Enumeration of symmetrized $SO(3)$-invariants}
The number of symmetrized $SO(3)$-invariants is expressed by formulas similar
to the one given for $G$. For example, if $\ell_1=\ell_2=\ell_3=\ell_4=\ell$, and if 
the rotations are defined by an axis $\bfn$ and an angle $\omega$, then the character 
of the rotation is $\chi_\ell(\omega)=\sin\big((2\ell+1)\omega/2\big)/\sin(\omega/2)$
and~\cite{VMK}
\begin{eqnarray*}
n^{S_1} &=& \frac{1}{8\pi}\int_0^{2\pi} d\omega \, \sin^2(\omega/2) 
\big(\chi_\ell(\omega)^4+ 3 \chi_\ell(2\omega)^2 \\
&& \qquad\qquad + 2\chi_\ell(2\omega)\chi_\ell(\omega)^2 + 
2\chi_\ell(4\omega)\big) = \ell+1.
\end{eqnarray*}
Similarly
\begin{eqnarray*}
n^{S_2} &=& \min(\ell_\alpha+1,\ell_\beta+1),\\
n^{S_3} &=& \min(2\ell_\alpha+1,2\ell_\beta+1),\\
n^{S_4} &=& \frac12 a^{\ell_\beta\ell_\gamma}(0)
- \frac12 a^{\ell_\beta\ell_\gamma}(2\ell_\alpha+1)
+ \frac12 b^{\ell_\beta\ell_\gamma}(2\ell_\alpha)
\\
n^{S_5} &=& a^{\ell_\alpha\ell_\beta}(|\ell_\gamma-\ell_\delta|)
-a^{\ell_\alpha\ell_\beta}(\ell_\gamma+\ell_\delta+1),
\end{eqnarray*}
where
\begin{eqnarray*}
a^{\ell\ell'}(m) &=& 
\begin{cases}
  2\min(\ell,\ell')+1 & \text{if }  
    |m| \le |\ell-\ell'|,\\
  \ell+\ell'-|m|+1 &\text{if }  |\ell-\ell'| \le |m| 
\le \ell+\ell',\\
 0 &\text{if }   |m| > \ell+\ell',
\end{cases}
\end{eqnarray*}
and
\begin{eqnarray*}
b^{\ell\ell'}(m) &=&
\begin{cases}
1 &\text{if } \ell+\ell'\text{ is even and } |m|\ge|\ell-\ell'|,\\
-1 &\text{if } \ell+\ell'\text{ is odd and } |m| > \ell+\ell',\\
0&\text{ otherwise}.
\end{cases}
\end{eqnarray*}

\subsection{Subduction from \texorpdfstring{$\calG$ to $G$}{G to H}}
We now come back to a general point group $G$, but for notational convenience 
we keep labeling the irreps of the larger group 
$\calG$ by $\ell_1, \ell_2, \ldots$,
as for $SO(3)$. However, we insist that the following formulas 
do not require $\calG=SO(3)$.

When lowering the symmetry, each irrep of the larger group branches 
into several irreps of the subgroup. For instance, the $\ell=2$ 
representation of $\calG=SO(3)$ splits into the representations $E_g$ 
and $T_{2g}$ of  $G=O_h$. Let us denote $\ell_1 \alpha$ the irrep 
$\alpha$ of $G$ that comes from $\ell_1$ of $\calG$, and idem for
$\ell_2 \beta$, $\ell_3 \gamma$, $\ell_4 \delta$ and $\ell \eta$.
The basis of the irreps of $\calG$, $\{| \ell m \rangle \}$ with
$1 \leq m \leq \dim \ell$, spans the same space as 
$\{ | \ell \alpha a \rangle \}$, where $\alpha$ runs over all 
the irreps subduced from $\ell$ and $1 \leq a \leq \dim \alpha$. 
As a consequence, the interaction elements can be labelled by 
$U^{(\ell_1 \ell_2 \ell_3 \ell_4)}_{\alpha a \beta b \gamma c \delta d}$ 
or $U^{(\ell_1 \alpha \ell_2 \beta \ell_3 \gamma \ell_4 \delta)}_{abcd}$ 
indiscriminately.

\subsubsection{Isoscalar factors}
 We showed in Sec.~\ref{sec:definvariant} that the group invariants 
 are basis-independent, but they depend on the point group.
In a $\calG$-symmetric point group, the interaction elements can be
expressed either in terms of $\calG$-invariants $I^{(\ell_1 \ell_2 \ell_3 \ell_4, l)}$
\begin{eqnarray}
&& U^{(\ell_1 \ell_2 \ell_3 \ell_4)}_{\alpha a \beta b \gamma c \delta d} \nonumber \\
 &&= \sum_{\ell \eta e} (\ell_1\alpha a \ell_2\beta b | \ell\eta e) 
 (\ell_3\gamma c \ell_4\delta d|\ell\eta e)^\ast I^{(\ell_1 \ell_2 \ell_3 \ell_4,\ell)}, \nonumber \\
\label{eq:UonG}
\end{eqnarray}
or in terms of the subduced $G$ invariants
$I^{(\ell_1 \alpha \ell_2  \beta \ell_3 \gamma \ell_4 \delta, \eta)}$
\begin{eqnarray}
&& U^{(\ell_1\alpha\ell_2\beta\ell_3\gamma\ell_4\delta)}_{abcd} \nonumber \\
&& \quad\quad =  \sum_{\eta e} (\alpha a \beta b | \eta e) 
(\gamma c \delta d | \eta e)^\ast I^{(\ell_1 \alpha \ell_2 \beta \ell_3 \gamma \ell_4 \delta, \eta)}. \nonumber \\
\label{eq:UonH}
\end{eqnarray}

Now, the Racah factorization lemma~\cite{Butler} states 
that the Clebsch-Gordan coefficients of $\calG$,
$(\ell_1\alpha a \ell_2\beta b|\ell\eta e)$, and those of 
$G$, $(\alpha a\beta b|\eta e)$, are related via complex numbers
$\isoscalar{\ell_1}{\ell_2}{\ell}{\alpha}{\beta}{\eta}$ called isoscalar factors
\begin{eqnarray}
(\ell_1\alpha a \ell_2\beta b|\ell\eta e) &=& 
\isoscalar{\ell_1}{\ell_2}{\ell}{\alpha}{\beta}{\eta} (\alpha a\beta b|\eta e).
\end{eqnarray}
Isoscalar factors are fundamental ingredients of group-subgroup 
symmetry caculations~\cite{Butler-75,Kibler-76,Kibler-77,
Butler,Piepho,Kibler-83}. They satisfy orthogonality 
relations~\cite{Butler}
\begin{eqnarray*}
\sum_{\ell}
\isoscalar{\ell_1}{\ell_2}{\ell}{\alpha}{\beta}{\eta}
\isoscalar{\ell_1}{\ell_2}{\ell}{\alpha'}{\beta'}{\eta}^*
&=& \delta_{\alpha\alpha'}\delta_{\beta\beta'},\\
\sum_{\alpha\beta}
\isoscalar{\ell_1}{\ell_2}{\ell}{\alpha}{\beta}{\eta}
\isoscalar{\ell_1}{\ell_2}{\ell'}{\alpha}{\beta}{\eta}^*
&=& \delta_{\ell\ell'}.
\end{eqnarray*}
We implicitely assumed that $\alpha, \beta, \eta, \ldots$ 
appear in the subduction of $\ell_1, \ell_2, \ell, \ldots$, respectively, 
and that $\eta$ belongs to the Clebsch-Gordan expansion of 
$\alpha \otimes \beta$ and $\alpha' \otimes \beta'$.
Isoscalar factors are usually calculated from Clebsch-Gordan 
coefficients, but their squares can be calculated from 
characters~\cite{Kibler-77,Prasad-79}.

By comparing the right-hand side of Eqs. \eqref{eq:UonG} and 
\eqref{eq:UonH} and using the usual orthogonality 
relations of Clebsch-Gordan coefficients, we get
\begin{equation}
I^{(\ell_1\alpha\ell_2\beta\ell_3\gamma\ell_4\delta,\eta)} = \sum_{\ell}
\isoscalar{\ell_1}{\ell_2}{\ell}{\alpha}{\beta}{\eta}
\isoscalar{\ell_3}{\ell_4}{\ell}{\gamma}{\delta}{\eta}^* 
 I^{(\ell_1 \ell_2 \ell_3 \ell_4,\ell)}
 \label{eq:subduction}
\end{equation}

To summarize the result of this section, 
$I^{(\ell_1\alpha \ell_2\beta \ell_3\gamma \ell_4\delta, \eta)}$ 
are $G$-invariants obtained for a system with a symmetry
group $\calG \supset G$. They can be directly compared to the 
$G$-invariants of a system with the actual
$G$ symmetry
group. Equation~\eqref{eq:subduction} can also be used to fit a 
set of $\calG$-invariants $I^{(\ell_1 \ell_2 \ell_3 \ell_4,\ell)}$
onto the $G$-invariants $I^{(\alpha\beta\gamma\delta,\eta)}$;
the mean squared error of the fit would measure the deviation 
of the lower-symmetry system from the one with the higher symmetry.

\subsection{Subduction from $O(3)$}
As explained in the introduction of this section, 
the irreps of $O(3)$ are the direct product
$L=(\ell,\epsilon_L)$ of an irrep $\ell$ of $SO(3)$
and an irrep $\epsilon_L=\pm1$ of $C_i$~\cite{LudwigFalter}.
The commutative group $C_i$ is  of order 2 
with two irreps denoted by $\epsilon_L=\pm1$
that are one-dimensional and
satisfy $\mathcal{I}\triangleright \varphi_L=\epsilon_L\varphi_L$.
Moreover, $\mathcal{I}$ commutes with $SO(3)$.
In general, the action of $\mathcal{I}$ on a tensor
product of irreps of $O(3)$ is 
\begin{eqnarray*}
\mathcal{I}\triangleright (\varphi_{L_1} \otimes \varphi_{L_2})
&=& \epsilon_{L_1}\epsilon_{L_2}(\varphi_{L_1} \otimes \varphi_{L_2}).
\end{eqnarray*}

The Clebsch-Gordan coefficients for $O(3)$
are
\begin{eqnarray*}
(L_1 m_1 L_2 m_2|L_3m_3) &=&
\delta_{\epsilon_{L_1}\epsilon_{L_2},\epsilon_{L_3}}
(\ell_1 m_1 \ell_2 m_2|\ell_3m_3).
\end{eqnarray*}
The $O(3)$ invariants are 
\begin{eqnarray*}
I^{(L_1L_2L_3L_4,L)} &=&
\delta_{\epsilon_{L_1}\epsilon_{L_2},\epsilon_{L}}
\delta_{\epsilon_{L_3}\epsilon_{L_4},\epsilon_{L}}
I^{(\ell_1\ell_2\ell_3\ell_4,\ell)}.
\end{eqnarray*}
The condition 
$\epsilon_{L_1}\epsilon_{L_2}=\epsilon_{L}=\epsilon_{L_3}\epsilon_{L_4}$
implies 
$\epsilon_{L_1}\epsilon_{L_2}\epsilon_{L_3}\epsilon_{L_4}=1$,
which is symmetric under permutation of 
$L_1,L_2,L_3,L_4$ and 
also implies
$\epsilon_{L_i}\epsilon_{L_j}=\epsilon_{L_k}\epsilon_{L_l}$,
where $(i,j,k,l)$ is any permutation of $(1,2,3,4)$.
Therefore,
$\langle I^{L_1L_2L_3L_4,L}\rangle$
is obtained from the same formula as 
$\langle I^{(\ell_1\ell_2\ell_3\ell_4,\ell)}\rangle$
up to the fact that the pair-conserving term
gets the factor 
$\delta_{\epsilon_{L_1}\epsilon_{L_2},\epsilon_{L}}
\delta_{\epsilon_{L_3}\epsilon_{L_4},\epsilon_{L}}$
while the pair-non-conserving term gets the factor
$\delta_{\epsilon_{L_1}\epsilon_{L_4},\epsilon_{L'}}
\delta_{\epsilon_{L_2}\epsilon_{L_3},\epsilon_{L'}}$.

In the calculations of the previous sections, 
we used spherical harmonics, for which 
$L=(\ell,\epsilon_L) $
with $\epsilon_L=(-1)^\ell$. 
Thus, in the following we denote by 
$\ell$ the $O(3)$ irrep $(\ell,(-1)^\ell)$.
Then
\begin{eqnarray*}
\mathcal{I}\triangleright U^{(\ell_1\ell_2\ell_3\ell_4)}_{m_1 m_2 m_3 m_4} &=&
U^{(\ell_1\ell_2\ell_3\ell_4)}_{m_1 m_2 m_3 m_4},
\end{eqnarray*}
because the selection rules for $3j$-symbols 
 imply 
$(-1)^{\ell_1+\ell_3}=(-1)^{\ell_2+\ell_4}$.
Similarly
$I^{(L_1L_2L_3L_4,L)}$
becomes 
$I^{(\ell_1\ell_2\ell_3\ell_4,L)}$
where 
$\epsilon_L=(-1)^{\ell_1+\ell_2}=(-1)^{\ell_3+\ell_4}$
because the $L$ in $I^{(\ell_1\ell_2\ell_3\ell_4,L)}$
does not correspond to a spherical harmonics,
for the same reason as the cross product of two 
vectors is a pseudovector (i.e. a ``vector'' which is even
under inversion).
Hence
\begin{eqnarray*}
\mathcal{I}\triangleright I^{(\ell_1\ell_2\ell_3\ell_4,L)}
&=& 
I^{(\ell_1\ell_2\ell_3\ell_4,L)}.
\end{eqnarray*}
Finally, since the condition 
$(-1)^{\ell_1+\ell_3+\ell_2+\ell_4}=1$ is invariant under permutation
of $L_1,L_2,L_3,L_4$, we also obtain
\begin{eqnarray*}
\mathcal{I}\triangleright \langle I^{(\ell_1\ell_2\ell_3\ell_4,L)}\rangle
&=& 
\langle I^{(\ell_1\ell_2\ell_3\ell_4,L)}\rangle,
\end{eqnarray*}
and $\langle I^{(\ell_1\ell_2\ell_3\ell_4,L)}\rangle$
is obtained from the same formula as
$\langle I^{(\ell_1\ell_2\ell_3\ell_4,\ell)}\rangle$,
except for the fact that the sum over $\ell'$ becomes
a sum over $L'=(\ell',\epsilon_{L'})$,
with $\epsilon_{L'}=(-1)^{\ell_1+\ell_4}=(-1)^{\ell_2+\ell_3}$,
the $6j$-symbols involving only $\ell$ and $\ell'$.

The subduction formulas are the same, provided we 
notice that, if $G$ contains $\mathcal{I}$, then \textit{gerade} irreps of $G$
can only arise from even $\ell_i$ and \textit{ungerade}
irreps from odd $\ell_i$.

\section{Conclusion}
\label{sec:concl}

Starting from the simple and familiar problem of calculating Coulomb integrals
in the most efficient way, we came to use surprisingly sophisticated
tools of group theory, such as Clebsch-Gordan coefficients, 
$6j$-symbols, corepresentation theory or Racah factorization
theorem, and had to recall the remarkable work by Derome and Sharp, which
was unjustly forgotten. These tools enabled us to provide explicit expressions 
for the Coulomb integrals in the most general case, i.e. for any orbital in any crystal
point group symmetry.  Moreover, instead of providing tables which would depend 
on the exact basis used for each irrep and on the phase choice of 
Clebsch-Gordan coefficients, we give here general and self-contained formulas.

Although  the spin degree of freedom was neglected in the present work, it is 
possible to take it into account as was done by Sugano and 
coauthors~\cite{TanabeSugano}, who considered Coulomb integrals 
between spin-1 and spin-0 states. This implies adding
the action of permutations
$(12)$ and $(34)$, for which the Coulomb integrals are odd for spin-1 
and even for spin-0. In other words, the full symmetric group $S_4$ should
be considered instead of $D_4$. The methods used in this paper can handle such a case, but
the formula for the permutation-symmetrized $G$-invariants would 
involve 24 terms instead of 8. 

It would also be tempting to refine the present treatment of
complex irreps, for example by
considering their real and imaginary parts and reducing the problem to the case of real representations. 
However, this would be a non-trivial extension of the present work,
because the resulting real representations would not be irreducible and
many of our proofs made a crucial use of Schur's lemma, which
holds only for irreps.
Pseudo-real irreps could possibly
be dealt with by using the fact that a representation and
its complex conjugate are related by a similarity transformation, generalizing what we
did for $SO(3)$.

Finally, another fruitful extension would be to deal with magnetic groups
and their corepresentations, which are used to describe the transport 
and response properties of magnetic and multiferroic 
materials~\cite{Kleiner-66,Butzal-82,Erb-20,Seemann-15,Yatsushiro-21}.
Although we dealt with corepresentations in the
present work, the theory of corepresentations is not as developed as the
theory of representations, and this extension would also be non-trivial.

\begin{acknowledgements}
C.L. acknowledges funding by the ``Fondation CFM pour la Recherche" via a 
\textit{Jean-Pierre Aguilar} PhD scholarship.
We thank Shintaro Hoshino for explaining certain technical aspects of his paper~\cite{Iimura-21}.
We are very grateful to Jean-Robert Derome for sending us a copy of his 
1965 PhD thesis~\cite{Derome-PhD}, without which the present work could not have been carried out.
\end{acknowledgements}


\begin{thebibliography}{99}
\bibitem{noteA}
We also assume that $U$ is real but this is not a loss of generality because any complex $U$ can be written $U=U_r+iU_i$ where both $U_r$ and $U_i$ are real.
\bibitem{note1}
The relation between our Clebsch-Gordan coefficients and the $3j$-symbols of Ref.~\cite{Derome-65} is 
$(\alpha a \beta b|\eta e) =  \sqrt{\dim\eta} \sum_{e'} (\alpha\beta\eta^*)^*_{abe'} (\eta)_{ee'}$, where
 $(\eta)_{ee'}$ is  the so-called $1j$-symbol.
\bibitem{note2}
The crucial point is the recoupling formula which we need in section~\ref{sec:perminv-real}.
In Butler's approach the recoupling formula involves $2jm$-symbols $\binom{\alpha\,\alpha^*}{a\,a'}$ which depend explicitly on the basis functions $\varphi^{(\alpha)}_a$. As a result, we cannot use them to identify recoupled $G$-invariants. On the contrary, Derome's recoupling formula makes use of $1j$-symbols $(\alpha)_{aa'}$, which are multiple of the identity for real and complex representations. Hence they simply add prefactors to the recoupled $G$-invariants.
\bibitem{note3}
Note that the formulas given in Ref.~\onlinecite{Derome-65} generalize the work by Sharp for 
quasi-ambivalent and multiplicity-free groups~\cite{Sharp}. All crystal point 
groups are quasi-ambivalent, see Ref.~\cite{Butler}.
\end{thebibliography}
\end{document}